\documentclass[11pt]{article}
\usepackage{graphicx}

\newcommand{\bra}{\langle}
\newcommand{\ket}{\rangle}
\newcommand{\R}{\mbox{\boldmath $ R $}}
\newcommand{\C}{\mbox{\boldmath $ C $}}
\newcommand{\Z}{\mbox{\boldmath $ Z $}}

\newcommand{\vect}[1]{\mbox{\boldmath $ #1 $}}

\def\dfrac#1#2{\displaystyle\frac{#1}{#2}}

\newcommand{\mapright}[2]
{\mathop{\hbox to 8mm{\rightarrowfill}}
\limits^{\scriptstyle #1}_{\scriptstyle #2}}

\makeatletter
\@addtoreset{equation}{section}
\makeatother
\renewcommand{\theequation}{\arabic{section}.\arabic{equation}}

\begin{document} 
\begin{flushright}
hep-th/0108208\\
KOBE-TH-01-04 \\
Ver.1 Aug 27, 2001 \\ 
Ver.2 Nov 22, 2001
\end{flushright}
\vspace*{6mm}
\begin{center}
{\Large \bf 
Spontaneous breaking of the C, P, and rotational symmetries
\vspace{3mm}\\
by topological defects
in two extra dimensions}\footnote{ 
To be published in Physical Review D.}
\vspace{8mm} \\
{\large
Makoto Sakamoto\footnote{e-mail: sakamoto@phys.sci.kobe-u.ac.jp},
\quad
Shogo Tanimura\footnote{e-mail: tanimura@kues.kyoto-u.ac.jp}
}
\vspace*{8mm} \\
{\small \it
${}^2 $Department of Physics, Kobe University, 
Rokkodai, Nada, Kobe 657-8501, Japan
\\
${}^3 $Department of Engineering Physics and Mechanics, 
Kyoto University \\ Kyoto 606-8501, Japan
}
\vspace{8mm}
\\
{\small Abstract}
\vspace{2mm}
\\
\begin{minipage}[t]{130mm}
\small
\baselineskip 5.6mm 
We formulate models of
complex scalar fields in the space-time 
that has 
a two-dimensional sphere as extra dimensions.
The two-sphere $ S^2 $ is assumed to have
the Dirac-Wu-Yang monopole as a background gauge field.
The nontrivial topology of the monopole
induces topological defects, i.e. vortices, in the scalar field.
When the radius of $ S^2 $ is larger than a critical radius,
the scalar field develops a vacuum expectation value 
and creates vortices in $ S^2 $.
Then the vortices break the rotational symmetry of $ S^2 $.
We exactly evaluate the critical radius as $ r_q = \sqrt{|q|}/\mu $,
where $ q $ is the monopole number and 
$ \mu $ is the imaginary mass of the scalar.
We show that the vortices repel each other.
We analyze the vacua of the models with one scalar field 
in each case of $ q=1/2, 1, 3/2 $
and find that:
when $ q=1/2 $, a single vortex exists;
when $ q=1   $, two vortices sit at diametrical points on $ S^2 $;
when $ q=3/2 $, three vortices sit at the vertices of the largest triangle 
on $ S^2 $.
The symmetry of the model
$ G = U(1) \times SU(2) \times CP $
is broken to
$ H_{1/2} = U(1)' $,
$ H_{1}   = U(1)'' \times CP $,
$ H_{3/2} = D_{3h} $,
respectively.
Here $ D_{3h} $ is the symmetry group of a regular triangle.
We extend our analysis to the doublet scalar fields
and show that the symmetry is broken 
from
$ G_{\mbox{\tiny doublet}} = U(1) \times SU(2) \times SU(2)_f \times P $
to
$ H_{\mbox{\tiny doublet}} = SU(2)' \times P $.
Finally we obtain the exact vacuum solution of
the model with the multiplet
$ (q_1, q_2, \cdots, q_{2j+1} ) = (j, j, \cdots, j ) $
and show that the symmetry is broken 
from
$ G_{\mbox{\tiny multiplet}} = U(1) \times SU(2) \times SU(2j+1)_f \times CP $
to
$ H_{\mbox{\tiny multiplet}} = SU(2)' \times CP' $.
Our results caution that
a careful analysis of dynamics of the topological defects is required for
construction of a reliable model that possesses such a defect structure.
\end{minipage}
\end{center}
\vspace{8mm}
\newpage
\baselineskip 5.3mm 

\renewcommand{\thefootnote}{\fnsymbol{footnote}}
\section{Introduction}
In spite of outstanding success of the Standard Model of particle physics 
in most of experimental tests,
there are several puzzles yet to be solved:
the neutrino oscillation \cite{neutrino exp},
the muon anomalous magnetic moment \cite{muon exp},
supersymmetry breaking,
the origin of the three generations,
the fermion mass hierarchy,
and the hierarchy problem of the electroweak, GUT, and Planck scales.
Recently,
the hypothesis that the space-time has 
compact extra dimensions \cite{Extra} 
is calling a renewed interest
since it provides possible solutions to these puzzles.
Here we cannot cover all the aspects of physics of extra dimensions.
Instead,
we concentrate on the aspects related 
to the origin of the generations
and to supersymmetry breaking.

The challenges to explain the fermion masses and generations 
by higher-dimensional models have a long history.
Hosotani \cite{Hosotani S1} found that 
dynamics of the Wilson loop in a multiply-connected space
breaks gauge symmetry and generates fermion masses.
This mechanism has been taken over to the string theory.
He also showed \cite{Hosotani S2} that the monopole background in $ S^2 $
generates a left-right asymmetric structure of fermions.
Akama, Rubakov, Shaposhnikov, and others,
in their early attempts \cite{Early},
proposed a hypothesis that
our four-dimensional space-time is a soliton-like object
embedded in the higher-dimensional space.
Recently, Arkani-Hamed and Schmaltz \cite{Arkani}
built a model
in that the Standard Model fields are confined to a thick wall 
in the extra dimensions,
and explained the hierarchy of Yukawa couplings.
Libanov and Troitsky \cite{Libanov} 
built a model
in that the fermions appear as zero modes trapped 
in the core of a topological defect in two extra dimensions, 
and attributed the number of the generations to 
the topological number of the vortex defect.
They with Fr{\`e}re \cite{Frere} also built another model, 
which has a single vortex with topological number one
but has three fermion generations.
Kawamura \cite{Kawamura}
suggested a mechanism of gauge symmetry breaking by the boundary condition in
the orbifold $ S^1 / \Z_2 $,
and recently Hall and Nomura \cite{Hall} 
completed a GUT model using this mechanism.
Bando et al.\cite{Bando} showed that
dynamics of Kaluza-Klein modes produces the hierarchy of Yukawa couplings
via the power-law behavior of the renormalized couplings to the energy scale.
Concerning dynamics in extra dimensions,
Randjbar-Daemi, Salam and Strathdee \cite{Salam:YM}
proved that
the $ SU(2) $-invariant solutions of the Yang-Mills system in 
$ \R^4 \times S^2 $ 
have tachyonic excitations and are unstable.
Recently, Dvali, Randjbar-Daemi and Tabbash \cite{Dvali} proposed a model
in $ \R^4 \times \C P^1 \times \C P^2 $
with the monopole and instanton background.
Using the mechanism proved previously,
they showed that
the extra-dimensional components of the gauge fields become tachyonic
and deduced 
the chiral fermion spectrum of the standard model.

On the other hand,
recently one of the authors \cite{Takenaga} proposed a new mechanism of 
supersymmetry breaking by a topologically nontrivial boundary condition
in one extra dimension $ S^1 $. 
In those models 
the translational symmetry in the direction of $ S^1 $ is spontaneously broken
and 
the supersymmetry is subsequently broken.
However, the model built on the one extra dimension is too simple 
to provide a realistic particle spectrum.
Therefore it is desirable to construct models on higher extra dimensions
and to analyze the patterns of symmetry breaking.

The purposes of this paper are threefold:
the first one is to construct models that have 
a two-dimensional sphere $ S^2 $ as extra dimensions;
the second is to analyze exhaustively the patterns 
of symmetries including the rotational symmetry of $ S^2 $;
the third one is to study the topological structure of the vacuum.

%
Topological defects in the vacuum break spatial symmetries.
For example, vortices in the sphere break the rotational symmetry.
Since the symmetry of the space governs the particle spectrum,
it is also worth studying.

To convey our basic idea
let us shortly review the model \cite{Sakamoto}
that exhibits
spontaneous breaking of the translational symmetry 
in one extra dimension. 
Assume that the space-time is 
a direct product of the Minkowski space with a circle, $ \R^n \times S^1 $,
which is equipped with the coordinate system 
$ ( x^0, x^1, \cdots, x^{n-1}, \phi ) $,
and in which the points 
$ \{ \phi + 2 \pi m \}_{m=0, \pm 1, \pm 2, \cdots} $
are identified.
The radius of $ S^1 $ is denoted by $ r $.
The model consists of a complex scalar field $ f(x,\phi) $ 
and has the Lagrangian
\begin{equation}
	{\cal L} =
	g^{\mu \nu} 
	\frac{\partial f^*}{\partial x^\mu}
	\frac{\partial f}{\partial x^\nu}
	-
	\frac{1}{r^2}
	\frac{\partial f^*}{\partial \phi}
	\frac{\partial f}{\partial \phi}
	+
	\mu^2 f^* f
	-
	\lambda (f^* f)^2.
	\label{action in S1}
\end{equation}
This model is invariant under the translations along $ S^1 $,
\begin{equation}
	e^{-iPd} : f(x,\phi) \mapsto f(x,\phi-d),
	\label{translation in S1}
\end{equation}
and also under the phase transformations,
\begin{equation}
	e^{-iQt} : f(x,\phi) \mapsto e^{-it} f(x,\phi).
	\label{phase in S1}
\end{equation}
Usually one imposes the periodic boundary condition on the field as
$ f (x,\phi + 2 \pi) = f(x,\phi) $,
but actually there is no a priori reason to impose it
if the field $ f $ itself is not a direct observable.
Instead we may impose the {\it twisted boundary condition}
\begin{equation}
	f (x,\phi + 2 \pi) = e^{- 2 \pi i \alpha} f(x,\phi)
	\label{twist}
\end{equation}
with the real parameter $ \alpha $ $ ( |\alpha| \le 1/2 ) $.
It should be noted that
the condition (\ref{twist}) is compatible with both the symmetries,
(\ref{translation in S1}) and (\ref{phase in S1}).
So the symmetries are not explicitly broken 
even when the twisted boundary condition is imposed.
The vacuum configuration
\begin{equation}
	\bra f(x,\phi) \ket = v \, e^{- i \alpha \phi}
	\label{vacuum in S1}
\end{equation}
minimizes the energy functional
\begin{equation}
	E =
	\int_0^{2 \pi} d \phi \, r
	\left\{
		\frac{1}{r^2}
		\frac{\partial f^*}{\partial \phi}
		\frac{\partial f}{\partial \phi}
		- \mu^2 f^* f
		+ \lambda (f^* f)^2
	\right\}
	=
	2 \pi r
	\left\{
		\frac{1}{r^2}  \alpha^2 v^2
		- \mu^2 v^2
		+ \lambda v^4
	\right\}.
	\label{energy in S1}
\end{equation}
Then, if we put $ r_c := | \alpha | / \mu $,
the minimum of $ E $ is realized by
\begin{equation}
	v^2 = 
	\left\{
		\begin{array}{ll}
		    0 
		    & \quad \mbox{for} \quad r \leq r_c,
		    \\
		    \bigg(
		    	1 - \dfrac{\alpha^2}{\mu^2 r^2}
		    \bigg) 
		    \dfrac{\mu^2}{2 \lambda} 
		    & \quad \mbox{for} \quad r > r_c.
		\end{array} 
	\right.
        \label{VEV for S1}
\end{equation}
We can see that
when the radius $ r $ of the circle is larger than
the critical radius $ r_c $,
the field $ f $ develops the non-vanishing non-constant 
vacuum expectation value (\ref{vacuum in S1})
and accordingly the translational symmetry (\ref{translation in S1}) is 
spontaneously broken.
However, the vacuum (\ref{vacuum in S1}) is still invariant under
the transformations given by $ e^{-i ( P + \alpha Q ) d } $.
Therefore the symmetry is broken from
$ U(1) \times U(1) $ to $ U(1) $.
The existence of the finite critical radius is a remarkable feature 
of this model.

Spontaneous breaking of translational symmetry has various physical implications;
for example, it can provide a new mechanism of supersymmetry breaking
as has been pointed out in the previous papers \cite{Takenaga}.
Because supercharges generate translations as
\begin{equation}
	\{ Q_\alpha, \bar{Q}_{\dot{\beta}} \} 
	= 2 \sigma_{\alpha \dot{\beta}}^\mu P_\mu,
	\label{SUSY algebra}
\end{equation}
breaking of the translational symmetry inevitably implicates
breaking of the supersymmetry.
More concretely, 
the super-transformation of a fermionic field is given by
\begin{equation}
	\delta_\xi \psi(x)
	=
	i \sqrt{2} \, \sigma^\mu \bar{\xi} \, \partial_\mu \varphi(x)
	+ \sqrt{2} \, \xi \, F(x).
	\label{SUSY transformation}
\end{equation}
The supersymmetry is broken if $ \bra F(x) \ket \ne 0 $.
This is a usual pattern of supersymmetry breaking
and is called $ F $-term breaking.
But the supersymmetry can be broken 
also if $ \partial_\mu \bra \varphi(x) \ket \ne 0 $.
The non-vanishing expectation value of the derivative
$ \partial_\mu \bra \varphi(x) \ket $
implies breaking of the translational symmetry.
Thus the translational symmetry breaking involves the supersymmetry breaking.
One of the authors \cite{Takenaga} constructed and analyzed
concrete models that
give rise to supersymmetry breaking 
via the twisted boundary condition in $ S^1 $.

The model mentioned above 
consists of only one complex scalar field in $ \R^n \times S^1 $
and has the $ U(1) $ internal symmetry.
Actually it is possible 
to construct models that have more fields and larger symmetries.
One of the authors \cite{Sakamoto} constructed 
a class of models that have real $ n $-component fields and $ O(n) $ symmetries
and exhaustively studied the patterns of breaking 
of the translational symmetry and of the $ O(n) $ symmetries.
However, 
models built
on the space-time with higher extra dimensions than $ S^1 $
are more desirable for application to particle physics,
because
higher dimensional manifolds can involve larger symmetries 
and richer particle spectra, 
which are useful for construction of realistic models.

As a step to the exploration to higher extra dimensions,
in the recent paper \cite{PL} we defined and studied a model 
that has the two-dimensional sphere $ S^2 $.
Then
the translational symmetry (\ref{translation in S1}) of $ S^1 $ is naturally
replaced by the rotational symmetry of $ S^2 $.
Moreover,
the twisted boundary condition (\ref{twist}) in $ S^1 $
is replace by the twisted patching condition in $ S^2 $, 
which is described as follows:
the spherical coordinate of $ S^2 $ is denoted by $ (\theta, \phi) $.
Let us introduce a pair of complex scalar fields, $ f_+ $ and $ f_- $,
which are defined 
in the upper $ ( 0 \le \theta < \pi ) $
and the lower $ ( 0 < \theta \le \pi ) $
region of $ S^2 $, respectively.
In the overlap of the two regions
we impose the {\it twisted patching condition}
\begin{equation}
        f_- (\theta, \phi) = e^{ - i m \phi } f_+ (\theta, \phi),
        \qquad
        0 < \theta < \pi,
        \label{patching condition}
\end{equation}
on the fields. 
Here $ m $ is an integer.
Eq. (\ref{patching condition}) is nothing but the gauge transformation
that accompanies the Dirac-Wu-Yang monopole 
$ A_\pm = (1 \mp \cos \theta) d \phi $.
In the body of this paper $ m $ is written as $ m = 2q $
and $ q $ is called the monopole charge.
When $ m \ne 0 $,
the vacuum expectation value $ \bra f_\pm (\theta,\phi) \ket $
cannot be a non-vanishing constant over $ S^2 $
owing to the patching condition.
Even if $ \bra f_\pm (\theta,\phi) \ket $ is nonzero in some region,
it should be zero at some points in $ S^2 $
for the topological obstruction.
The zero points of $ \bra f_\pm (\theta,\phi) \ket $ are called vortices.
Thus the rotational symmetry is pinned down by the vortices.

In the previous paper \cite{PL}
we studied in detail the model 
that has a complex scalar field in the monopole background
in $ \R^n \times S^2 $.
There we proved the existence of a critical radius of $ S^2 $;
when the radius of $ S^2 $ exceeds the critical radius,
the field develops vortices and breaks the rotational symmetry.
We also estimated the exact value of the critical radius.
But our treatment was restricted to the model
that has only one complex scalar field.
Besides, discrete symmetries including $ C $ and $ P $
were missed from our consideration.

In this paper
we first examine the symmetries of the model 
of one complex scalar field in $ \R^n \times S^2 $ with the monopole background
in detail.
We then find that the model has a $ \Z_2 $ symmetry,
which is analogous to the $ CP $ symmetry.
We show that
the rotational and the discrete symmetries are spontaneously broken 
when the radius of $ S^2 $ is larger than the critical radius.
We calculate the locations of the vortices of the stable vacuum.
We also clarify the structure of Nambu-Goldstone bosons 
that are associated with the symmetry breaking.
Furthermore we extend our argument to models 
that have more scalar fields and larger symmetries.
If the matter multiplet has the charge 
$ (q_1, q_2, \cdots, q_n ) $ and if $ \sum_i q_i = 0 $,
then the monopole gauge field can be embedded in
an $ SU(n) $ gauge field, which is free from the twisted patching condition.
Moreover,
when $ (q_1, q_2, \cdots, q_{2j+1} ) = (j, j, \cdots, j ) $,
we obtain the exact vacuum and 
show that a modified rotational symmetry is left unbroken.
Thus the patterns of symmetry breaking strongly depend on 
the matter contents of the model.
Although our study is motivated 
by an attempt to get a new mechanism of supersymmetry breaking,
our model is not yet made supersymmetric,
or does not exhibit supersymmetry breaking, either.
We should declare that
our study is restricted to bosonic fields at the present stage.

This paper is organized as follows:
in Section 2 we define the model
of one complex scalar field in $ \R^n \times S^2 $ 
and characterize its complete symmetry as $ G = U(1) \times SU(2) \times CP $.
We find that the model is not invariant under $ C $ and $ P $ separately
but it is actually invariant under the combined $ CP $ transformation.
We also estimate the exact critical radius as 
$ r_q = \sqrt{|q|} / \mu $,
where 
$ q $ is the monopole number and 
$ \mu $ is the imaginary mass of the scalar.
Although this part is a repetition of the previous paper \cite{PL},
we include it here to make the present paper self-contained.
We calculate the approximate vacuum 
in each case of $ q =1/2, 1, 3/2 $ concretely
and find that:
when $ q=1/2 $, a single vortex exists;
when $ q=1   $, two vortices sit at opposite two points on $ S^2 $;
when $ q=3/2 $, three vortices sit at the vertices of the largest triangle 
on $ S^2 $.
Then we observe that the symmetry is broken to
$ H_{1/2} = U(1)' $,
$ H_{1}   = U(1)'' \times CP $,
$ H_{3/2} = D_{3h} $ for each case.
Here $ D_{3h} $ is the symmetry of a regular triangle 
including the reflection transformation.
Since our approximate calculation is based on the variational method,
we give an argument to justify the method.

In Section 3 we formulate the model of doublet fields
with monopole charges $ (q_1,q_2) = (q, -q) $.
Then we show that the scalar and the gauge fields are embedded in $ SU(2) $
and are transformed into fields that are single-valued over the whole $ S^2 $.
The model has the symmetry
$ G_{\mbox{\tiny doublet}} = U(1) \times SU(2) \times SU(2)_f \times P $.
Charge conjugation is included in $ SU(2)_f $.
Then we obtain the exact vacuum of the $ q=1/2 $ doublet model
and show that the symmetry is broken to
$ H_{\mbox{\tiny 1/2 doublet}} = SU(2)' \times P $
when $ r > (\sqrt{2} \mu)^{-1} $.
In this model vortices do not appear and 
a modified rotational symmetry remains unbroken.

In Section 4 we consider models that consist of arbitrary multiplet fields with
$ (q_1, q_2, \cdots, q_n ) $.
Then we prove that the fields become free from the Dirac singularity
if $ \sum_i q_i = 0 $.
We show that
a specific model with the charge multiplet
$ (q_1, q_2, \cdots, q_{2j+1} ) = (j, j, \cdots, j ) $ has the symmetry
$ G_{\mbox{\tiny multiplet}} = 
U(1) \times SU(2) \times SU(2j+1)_f \times CP $.
We obtain its exact vacuum solution and show that the symmetry is broken to
$ H_{\mbox{\tiny multiplet}} = SU(2)' \times CP' $
when $ r > \sqrt{j}/\mu $.

In conclusion we speculate about
generalizations and applications of this work.

\section{Singlet models}
\subsection{Definitions and symmetries}
{}First, we define a model, which exhibits
spontaneous breaking of the rotational symmetry.
Our model is defined in the space-time $ \R^n \times S^2 $,
where 
$ \R^n $ is an $ n $-dimensional Minkowski space
and 
$ S^2 $ is a two-dimensional sphere of the radius $ r $.
The Cartesian coordinate of $ \R^n $ is denoted by 
$ (x^0,x^1, \cdots, x^{n-1} ) $
while 
the spherical coordinate of $ S^2 $ is denoted by $ (\theta,\phi) $.
The space $ \R^n $ is equipped with
the metric $ g_{\mu \nu} = \mbox{diag} (+1,-1,\cdots,-1) $.
Our model consists of a complex scalar field $ f $ in $ \R^n \times S^2 $
with a background gauge field $ A $ in $ S^2 $.
The gauge field $ A $ is fixed to be the Dirac-Wu-Yang monopole \cite{Wu},
which is defined as follows:
two open sets of $ S^2 $,
$ U_+ = \{ (\theta,\phi) | \theta \ne \pi \} $ and
$ U_- = \{ (\theta,\phi) | \theta \ne 0   \} $,
cover $ S^2 = U_+ \cup U_- $.
The monopole field $ A $ is described by the pair of 1-forms $ (A_+,A_-) $,
\begin{equation}
        A_{\pm}(\theta,\phi) = ( \pm 1 - \cos \theta ) d \phi 
        \quad \mbox{in} \; U_{\pm}.
        \label{gauge}
\end{equation}
The associated magnetic field is 
$ B = dA_+ = dA_- = \sin \theta \, d \theta \wedge d \phi $.
Then the total magnetic flux is given by 
$ \int B = 4 \pi $.
The complex scalar field $ f $ is described by a pair of fields $ (f_+, f_- ) $,
where $ f_\pm $ is a smooth function over $ U_\pm $, respectively.
The covariant derivative of $ f $ is defined by
\begin{equation}
	Df_\pm = df_\pm - i q A_\pm f_\pm 
	\quad \mbox{in} \; U_{\pm}
	\label{covariant}
\end{equation}
with the coupling constant $ q $.
It is represented as the product $ q = eg $
of the electric charge $ e $
and the magnetic charge $ g $
that is usually defined by the magnetic flux $ \int B = 4 \pi g $.
However, 
we conveniently make the constant $ q $ absorb both $ e $ and $ g $ 
into its definition
and 
we simply call $ q $ the magnetic number or the magnetic charge.
The pairs of the fields,
$ (A_+, A_-) $ and $ (f_+,f_-) $, are patched together 
by the gauge transformation \cite{Wu},
\begin{equation}
        A_- = A_+ - 2 \, d \phi,
        \qquad
        f_- = e^{ - 2 i q \phi } f_+ 
	\qquad \mbox{in} \; U_+ \cap U_-.
        \label{patch}
\end{equation}
Because the fields $ f_\pm $ must each be single-valued,
$ 2q $ must be an integer.
The action of our model is
\begin{eqnarray}
        S_A [f] &=&
        \int d^n x\, d \theta d \phi \, r^2 \sin \theta   
        \Bigg\{
                g^{\mu \nu}
                \frac{\partial f_{\pm}^*}{\partial x^\mu}
                \frac{\partial f_{\pm}  }{\partial x^\nu}
                - \frac{1}{r^2}
                \Bigg|
                        \frac{\partial f_{\pm}}{\partial \theta}
                \Bigg|^2
	\nonumber \\ && 
                -
                \frac{1}{r^2 \sin^2 \theta}
                \Bigg|
                        \frac{\partial f_{\pm}}{\partial \phi}
                        - i q
                        ( \pm 1 - \cos \theta ) f_{\pm}
                \Bigg|^2
                + \mu^2 f_{\pm}^* \! f_{\pm}
				- \lambda (f_{\pm}^* \! f_{\pm})^2
        \Bigg\},
        \label{action}
\end{eqnarray}
where $ \mu^2, \lambda > 0 $ are real parameters.

This model has a global symmetry 
\begin{equation}
	G =  U(1) \times SU(2) \times CP,
	\label{global symmetry}
\end{equation}
as being seen below.
The $ U(1) $ symmetry is defined as a family of the transformations
\begin{equation}
        e^{-iQt} :
        f_\pm  \; \mapsto \; e^{-iqt} f_\pm 
        \label{U(1)}
\end{equation}
with $ t \in \R $.
The generator of the transformation is $ Q = q $.
On the other hand,
the elements of $ SU(2) $ act on $ S^2 $ as rotation transformations.
Of course, the explicit form of the background gauge field 
$ A_\pm = ( \pm 1 - \cos \theta ) d \phi $
is not invariant under arbitrary rotations.
To leave it 
invariant,
a rotation transformation is to be complemented by a gauge transformation.
Such a gauge transformation can be calculated as follows:
the spherical coordinate $ (\theta,\phi) $ 
is assigned to a point $ p \in S^2 $.
We introduce two maps $ s_\pm : U_\pm \to SU(2) $ by
\begin{equation}
        s_\pm (p) :=
        e^{-i \sigma_3 \phi  /2}
        e^{-i \sigma_2 \theta/2}
        e^{\pm i \sigma_3 \phi  /2},
        \label{sections}
\end{equation}
where $ \{ \sigma_1, \sigma_2, \sigma_3 \} $ are the Pauli matrices.
Note that $ s_\pm (p) $ are smooth in $ U_\pm $, respectively.
These maps have the property
\begin{equation}
        s_\pm (p) \cdot \sigma_3 \cdot s_\pm (p)^{-1}
        =
        \sigma_1 \sin \theta \cos \phi +
        \sigma_2 \sin \theta \sin \phi +
        \sigma_3 \cos \theta.
        \label{sections property}
\end{equation}
Because
\begin{eqnarray}
	s_\pm^{-1} d s_\pm 
& = &	- \frac{i}{2}
	\Bigl\{
		\sigma_1 
		( \mp \sin \phi \, d \theta
		- \cos \phi \sin \theta \, d \phi)
		+
		\sigma_2 
		( \cos \phi \, d \theta
		\mp \sin \phi \sin \theta \, d \phi )
	\nonumber \\
&&
	\qquad
		+
		\sigma_3 
		( \mp 1 + \cos \theta ) d \phi 
	\Bigr\},
	\label{Maurer-Cartan form}
\end{eqnarray}
we get
\begin{equation}
	A_\pm 
	= -i \, \mbox{tr} ( \sigma_3 \cdot s_\pm^{-1} d s_\pm ).
	\label{MC gives A}
\end{equation}
Let indices $ \alpha $ and $ \beta $ denote either $ + $ or $ - $.
Suppose that a point 
$ p \in U_\alpha $ is transformed to
$ g^{-1} p \in U_\beta $ by an element $ g \in SU(2) $.
Then we define an element of $ SU(2) $ by
\begin{equation}
        W_{\alpha \beta} (g;p) :=
        s_\alpha(p)^{-1} \cdot g \cdot s_\beta(g^{-1} p),
        \label{Wigner}
\end{equation}
which is called the Wigner rotation in the context of the representation theory,
for example, in \cite{Wigner} and \cite{Landsman}.
Since 
$ W_{\alpha \beta} \cdot \sigma_3 \cdot W_{\alpha \beta}^{-1} = \sigma_3 $,
the value of the Wigner rotation has a form 
\begin{equation}
        W_{\alpha \beta} (g;p) =
        e^{-i \sigma_3 \omega /2}
        \label{Wigner in U(1)}
\end{equation}
with a real number $ \omega $,
which defines a function $ \omega_{\alpha \beta} (g;p) $.
Since Eq. (\ref{Wigner}) implies
\begin{equation}
	s_\beta(g^{-1} p) =
	g^{-1} \cdot 
	s_\alpha(p) \cdot 
	W_{\alpha \beta} (g;p),
	\label{Wigner deformed}
\end{equation}
the gauge field rotated by $ g \in SU(2) $ becomes
\begin{eqnarray}
	A_\beta (g^{-1}p)
& = &
	-i \, \mbox{tr} ( \sigma_3 \cdot 
	s_\beta^{-1}(g^{-1}p) d s_\beta (g^{-1}p) )
	\nonumber \\
& = &
	-i \, \mbox{tr} 
	\Big( 
		\sigma_3 \cdot 
		W_{\alpha \beta}(p)^{-1} s_\alpha(p)^{-1} g
		\, d 
		( g^{-1} s_\alpha(p) W_{\alpha \beta}(p) )
	\Big)
	\nonumber \\
& = &
	-i \, \mbox{tr} 
	\Big( 
		\sigma_3 \cdot 
		s_\alpha(p)^{-1} \, d s_\alpha(p) 
	\Big)
	-i \, \mbox{tr} 
	\Big( 
		\sigma_3 \cdot 
		W_{\alpha \beta}(p)^{-1} \, d W_{\alpha \beta}(p) 
	\Big)
	\nonumber \\
& = &
	A_\alpha (p)
	-
        d \omega_{\alpha \beta} (p).
	\label{rotated A}
\end{eqnarray}
Therefore, 
a sequence of
the rotation by $ g \in SU(2) $ and 
the gauge transformation by $ \omega_{\alpha \beta} (g;p) $,
\begin{equation}
        \varphi(g) :
        f_\alpha(p) \mapsto 
        f'_\alpha (p) =
        e^{iq \omega_{\alpha \beta} (g; p)} f_\beta(g^{-1} p),
        \label{SU(2)f}
\end{equation}
leaves the monopole gauge field invariant as
\begin{equation}
        A_\alpha(p) 
        \mapsto
        A'_\alpha(p) =
        A_\beta(g^{-1}p) + \, d \omega_{\alpha \beta} (g;p)
        = A_\alpha(p).
        \label{SU(2)A} 
\end{equation}
Under this transformation
the action (\ref{action}) is also left invariant.
It is easily verified that the Wigner rotations satisfy
\begin{equation}
        W_{\alpha \beta} (g    ; p) 
        W_{\beta \gamma} (g'   ; g^{-1} p) =
        W_{\alpha \gamma} (g g'; p).
        \label{composition of Wigner}
\end{equation}
For the identity transformation $ e \in SU(2) $ we have
\begin{equation}
        W_{-+} (e; p) 
        = s_-(p)^{-1} \cdot s_+(p) 
        = e^{i \sigma_3 \phi},
        \label{identity Wigner}
\end{equation}
which implies
\begin{equation}
	\omega_{-+}(e;p) = - 2 \phi
	\label{identity}
\end{equation}
in the place of (\ref{Wigner in U(1)}).
Hence (\ref{SU(2)f}) reproduces the patching condition
$ f_- = e^{-2 i q \phi} f_+ $,
as it should be.

Although calculation of the concrete value of 
the Wigner rotation (\ref{Wigner}) is cumbersome,
to give a definite example we calculate it for 
$ g = e^{-i \sigma_3 \gamma/2} $,
which is a rotation around the $ z $-axis by the angle $ \gamma $.
The point $ p = (\theta, \phi) \in U_\pm $ is then transformed to
$ g^{-1} p = (\theta, \phi - \gamma) \in U_\pm $, 
and from the definitions (\ref{sections}) and (\ref{Wigner}) we get
\begin{equation}
        W_{\pm \pm} ( e^{-i \sigma_3 \gamma/2} ;p) 
        = 
        e^{\mp i \sigma_3 \gamma /2},
        \label{Wigner for z-rotation}
\end{equation}
which implies 
\begin{equation}
	\omega_{\pm \pm} (e^{-i \sigma_3 \gamma/2}; p) = \pm \gamma = 
	\mbox{constant}
	\label{z-Wigner}
\end{equation}
in (\ref{Wigner in U(1)}).
The transformation (\ref{SU(2)f}) now becomes
\begin{equation}
        \varphi (e^{-i \sigma_3 \gamma/2}) :
        f_\pm (\theta,\phi) 
	\mapsto
        e^{\pm iq \gamma} f_\pm (\theta,\phi - \gamma).
        \label{f to f'}
\end{equation}

{}For later use, let us calculate the Wigner rotation 
for the rotation around the $ x $-axis by the angle $ \pi $,
which is represented by
$ g = e^{-i \sigma_1 \pi/2} = -i \sigma_1 $.
The point $ p = (\theta, \phi) \in U_\pm $ is then transformed to
$ g^{-1} p = (\pi-\theta, -\phi) \in U_\mp $.
Then the corresponding Wigner rotation is calculated to be
\begin{equation}
	W_{\pm \mp} (e^{-i \sigma_1 \pi/2}; p) 
	=
	-i \sigma_3 
	=
	e^{ -i \sigma_3 \pi/2 },
	\label{Wigner for x-rotation}
\end{equation}
and therefore we have 
\begin{equation}
	\omega_{\pm \mp}(e^{-i \sigma_1 \pi/2}; p) = \pi
	\label{x-pi-Wigner}
\end{equation}
in the place of (\ref{Wigner in U(1)}).
The transformation (\ref{SU(2)f}) now becomes
\begin{equation}
        \varphi (e^{-i \sigma_1 \pi/2}) :
        f_\pm (\theta,\phi) 
	\mapsto
        e^{iq \pi} f_\mp (\pi-\theta, -\phi).
        \label{f to f' x-axis}
\end{equation}

For later use
we would like to present infinitesimal transformations.
The generators $ \{ L_1, L_2, L_3 \} $ of the $ SU(2) $ transformations 
(\ref{SU(2)f}) are defined by
\begin{eqnarray}
	- i L_k f_\alpha (p)
	:=
	\left. 
		\frac{d}{d \gamma} 
		\varphi( e^{-i \sigma_k \gamma /2}) f_\alpha (p) 
	\right|_{\gamma=0}.
	\label{Ls}
\end{eqnarray}
Then their concrete forms are expressed as
\begin{eqnarray}
&&
	L_1 f_\pm = i
	\bigg[
		\sin \phi \frac{\partial}{\partial \theta}
		+
		\frac{\cos \phi}{\sin \theta}
		\Big(
			\cos \theta \frac{\partial}{\partial \phi}
			+ i q ( 1 \mp \cos \theta )
		\Big)
	\bigg] f_\pm,
	\label{L_1} \\
&&
	L_2 f_\pm = i
	\bigg[
		- \cos \phi \frac{\partial}{\partial \theta}
		+
		\frac{\sin \phi}{\sin \theta}
		\Big(
			\cos \theta \frac{\partial}{\partial \phi}
			+ i q ( 1 \mp \cos \theta )
		\Big)
	\bigg] f_\pm,
	\label{L_2} \\
&&
	L_3 f_\pm = i
	\bigg[
		- \frac{\partial}{\partial \phi} \pm i q
	\bigg] f_\pm.
	\label{L_3}
\end{eqnarray}
They satisfy $ [ L_j, L_k ] = i \epsilon_{jk\ell} L_\ell $
and exactly agree with the angular momentum operators 
given by Wu and Yang \cite{Wu}.
We make also the linear combinations
\begin{eqnarray}
	L_+ f_\pm
& := &
	( L_1 + i L_2 ) f_\pm
	\nonumber \\
& = &
	+ e^{ + i \phi }
	\bigg[
		\frac{\partial}{\partial \theta}
		+ 
		\frac{i}{\sin \theta}
		\Big(
			\cos \theta \frac{\partial}{\partial \phi}
			+ i q ( 1 \mp \cos \theta )
		\Big)
	\bigg] f_\pm,
	\label{L+} \\
	L_- f_\pm
& := &
	( L_1 - i L_2 ) f_\pm
	\nonumber \\
& = &
	- e^{ - i \phi }
	\bigg[
		\frac{\partial}{\partial \theta}
		- 
		\frac{i}{\sin \theta}
		\Big(
			\cos \theta \frac{\partial}{\partial \phi}
			+ i q ( 1 \mp \cos \theta )
		\Big)
	\bigg] f_\pm.
	\label{L-}
\end{eqnarray}

We should mention the discrete symmetry that our model possesses.
We can show that
although either parity $ P $ or charge conjugation $ C $ 
is not symmetry of the model,
their combination $ CP $ is its symmetry.

Let us define the parity transformation of our model.
Actually, parity transforms a model with the monopole number $ q $
to another model with $ -q $.
Parity is inversion of $ S^2 $ as
\begin{equation}
	P : S^2 \to S^2; \;
	( \theta, \phi )
	\mapsto
	( \pi -  \theta, \pi + \phi ).
	\label{parity in spherical}
\end{equation}
Parity changes appearance of the patching conditions (\ref{patch}) as
\begin{eqnarray}
	A_- ( \pi- \theta, \pi+\phi )
& = &	A_+ ( \pi-\theta, \pi+\phi ) - 2 \, d (\pi + \phi)
	\nonumber \\
& = &	A_+ ( \pi-\theta, \pi+\phi ) - 2 \, d \phi,
	\\
	f_- ( \pi-\theta, \pi+\phi )
& = &	e^{ - 2 i q (\pi + \phi) }      f_+ ( \pi-\theta, \pi+\phi )
	\nonumber \\
& = &	e^{ - 2 i q \phi} \, e^{-2 i q \pi} f_+ ( \pi-\theta, \pi+\phi ),
	\label{patch parity}
\end{eqnarray}
which can be rearranged as
\begin{eqnarray}
	- A_+ ( \pi-\theta, \pi+\phi )
	& = &
	- A_- ( \pi-\theta, \pi+\phi ) - 2 \, d \phi,
	\\
	e^{-i q \pi} f_+ ( \pi-\theta, \pi+\phi )
	& = &
	e^{ 2 i q \phi} \, 
	e^{ i q \pi} f_- ( \pi-\theta, \pi+\phi ).
	\label{paritied patch}
\end{eqnarray}
If we define the parity transformation of the fields as
\begin{eqnarray}
	&& \varphi_P : 
	A_\pm (\theta, \phi)
	\mapsto 
	- A_\mp (\pi - \theta, \pi + \phi),
	\label{parity A} \\
	&& \varphi_P : 
	f_\pm (\theta, \phi)
	\mapsto 
	e^{\pm i q \pi} f_\mp (\pi - \theta, \pi + \phi),
	\label{parity f}
\end{eqnarray}
it changes the monopole number from $ q $ to $ -q $
as seen by comparing (\ref{patch}) with (\ref{paritied patch}).
Besides, the monopole background field is left invariant as
\begin{eqnarray}
	A_\pm (\theta, \phi) \; \mapsto \;
	- A_{\mp} ( \pi - \theta, \pi + \phi )
	& = & - ( \mp 1 - \cos (\pi-\theta ) ) d ( \pi + \phi )
	\nonumber \\
	& = & ( \pm 1 - \cos \theta ) d \phi
	\nonumber \\
	& = & A_{\pm} ( \theta, \phi ).
	\label{sign change under parity}
\end{eqnarray}

Charge conjugation is defined simply by complex conjugation
\begin{equation}
	\varphi_C : 
	f_\pm \mapsto f_\pm^*.
	\label{charge f}
\end{equation}
Then the patching condition (\ref{patch}) becomes
\begin{equation}
	f_-^* = e^{ 2 i q \phi } f_+^*.
	\label{charge patch}
\end{equation}
Therefore charge conjugation is not symmetry of the model, either.
Instead it transforms a model with the monopole number $ q $
to another model with $ -q $.

Because the combined $ CP $ transformation 
turns the sign of the monopole number back to the original,
$ CP $ is symmetry of the model.
{}For later use we write down the $ CP $ transformation explicitly as
\begin{equation}
	\varphi_{CP} = 
	\varphi_C \circ \varphi_P : 
	f_\pm  ( \theta, \phi )
	\mapsto 
	e^{\mp i q \pi} f_\mp^* ( \pi - \theta, \pi + \phi ).
	\label{CP f}
\end{equation}

\subsection{Rotational symmetry breaking}
Assume that the monopole number $ q $ is not zero
and that the fields $ (f_+,f_-) $ are continuous functions.
Then it is proved that
if the field $ f $ is rotationally invariant, 
it must vanish identically as $ f \equiv 0 $ over $ S^2 $.
The contraposition says that
if $ f $ is not identically zero, 
$ f $ cannot be rotationally invariant over $ S^2 $.
Actually, even if $ f $ takes nonzero values in some region of $ S^2 $,
the value of $ f (\theta,\phi) $ 
must vanish at some points in $ S^2 $.
It can be shown that 
the number of the zero points of $ f $ is $ 2|q| $\footnote{
To be precise, what can be shown is that 
the sum of indices of all the zero points is equal to $ 2q $.
We can assign an integer index to each vortex by counting the winding number 
of the scalar field around the vortex with taking its sign into account.
Then we can assert that the sum of the indices is $ 2q $.
However, the number of vortices may appear more or less than $ 2|q| $
by pair creation of vortex and anti-vortex 
or by duplication of vortices.
}.
Hence the zero points pin down the rotational symmetry.
The zero points are called vortices \cite{vortex}.

Here we describe the outline of the proof of the theorem 
about the number of zero points.
The continuous function $ f_+ : U_+ \to \C $ defines
a family of loops
$ \{ \ell_+^{\,\theta} : S^1 \to \C $ $ ( 0 \le \theta \le \pi/2 ) \} $
by $ \ell_+^{\,\theta} (\phi) := f_+ (\theta,\phi) $.
Similarly, 
the continuous function $ f_- : U_- \to \C $ defines
a family of loops
$ \{ \ell_-^{\,\theta} : S^1 \to \C $ $ ( \pi/2 \le \theta \le \pi ) \} $
by $ \ell_-^{\,\theta} (\phi) := f_- (\theta,\phi) $.
Since $ f_\pm $ are single-valued over $ U_\pm $,
$ \ell_+^{\,0}   $ and 
$ \ell_-^{\,\pi} $ are shrunk loops.
Suppose that $ f_+ $ does not vanish in the upper hemisphere, 
$ 0 \le \theta \le \pi/2 $.
Then the loops 
$ \{ \ell_+^{\,\theta}(\phi) = f_+ (\theta,\phi) $
$ ( 0 \le \theta \le \pi/2 ) \} $
never touch nor cross the zero in the complex plane.
Therefore, the loop $ \ell_+^{\,\pi/2} $ does not wind around the zero.
Accordingly, 
$ \ell_-^{\,\pi/2} (\phi) 
= f_- (\pi/2, \phi)
= e^{-2iq \phi} f_+ (\pi/2, \phi) 
= e^{-2iq \phi} \ell_+^{\,\pi/2} (\phi) 
$
runs around the zero in the complex plane $ 2q $ times
in the clockwise direction.
In the limit $ \theta \to \pi $
the loop $ \ell_-^{\,\theta} $ 
continuously shrinks into a point in the complex plane.
Hence the loop $ \ell_-^{\,\theta} $ crosses the zero $ 2|q| $ times 
during the change of $ \theta $ over $ \pi/2 \le \theta \le \pi $.
Therefore $ f_-$ has $ 2|q| $ zero points.
We may relax the assumption and allow $ f_+ $ to have zero points in $ U_+ $.
We can similarly show that 
the total number of zero points of $ f_\pm $ is equal to $ 2|q| $.
Then the theorem about the number of zero points is proved.

Let us turn to the theorem about rotational symmetry breaking.
The statement that $ f $ is rotationally invariant means that
both $ f_+ $ and $ f_- $ remain invariant under the transformations 
(\ref{SU(2)f}) by $ SU(2)$.
Assume that $ f_\pm $ are rotationally invariant.
We have already shown that 
$ f_\pm $ vanish at some point in $ S^2 $ if $ q \ne 0 $.
Hence $ f_\pm $ should vanish everywhere
because the group $ SU(2) $ acts on $ S^2 $ transitively.
Accordingly,
the rotationally invariant $ f_\pm $ must be identically zero.
The proof is over.

The last theorem tells that 
if the scalar field exhibits a nonzero vacuum expectation value $ \bra f \ket $,
the rotational symmetry is necessarily broken.
Now we would like to examine a condition for rotational symmetry breaking.
In this paper we analyze the model only 
at the classical level.
Moreover, since the translational symmetry in $ \R^n $ is kept unbroken,
what we need to find is the vacuum configuration $ \bra f (\theta, \phi) \ket $ 
that minimizes the classical energy functional
\begin{equation}
        E 
	= 
        \int d \theta d \phi \, r^2 \sin \theta
        \Bigg\{
                \frac{1}{r^2}
                \bigg|
                        \frac{\partial f_\pm}{\partial \theta}
                \bigg|^2
                +
                \frac{1}{r^2 \sin^2 \theta}
                \bigg|
                        \frac{\partial f_\pm}{\partial \phi}
                        - iq
                        ( \pm 1 - \cos \theta ) f_\pm
                \bigg|^2
                - \mu^2 f_{\pm}^* \! f_{\pm}
				+ \lambda (f_{\pm}^* \! f_{\pm})^2
        \Bigg\}.
        \label{energy}
\end{equation}
The variation of the gradient energy with respect to $ f_\pm^* $ gives
the Laplacian coupled to the monopole,
\begin{equation}
        - \Delta_q f_\pm
:=
        - \left[
                \frac{1}{\sin \theta}
                \frac{\partial}{\partial \theta}
                \bigg(
                        \sin \theta
                        \frac{\partial}{\partial \theta}
                \bigg)
                +
                \frac{1}{\sin^2 \theta}
                \bigg(
                        \frac{\partial}{\partial \phi}
                        -iq ( \pm 1 - \cos \theta )
                \bigg)^2
        \right] f_\pm.
        \label{Laplacian}
\end{equation}
The eigenvalue problem of the monopole Laplacian was solved by 
Wu and Yang \cite{Wu}
and the solutions are summarized in a neat way by
Coleman \cite{monopole};
its eigenfunctions are expressed in terms of
the matrix elements of unitary representations
of $ SU(2) $,
$ D^j_{mq} (\theta, \phi, \psi) =
\bra j, m | 
e^{-i J_3 \phi} \, e^{-i J_2 \theta} \, e^{- i J_3 \psi} | j,q \ket $,
as 
\begin{equation}
        f_\pm (\theta,\phi)
        =
        D^j_{mq} (\theta, \phi, \mp \phi)
        =
        \bra j, m | 
        e^{-i J_3 \phi} \, e^{-i J_2 \theta} \, e^{\pm i J_3 \phi} 
        | j,q \ket
        =
        e^{-i (m \mp q) \phi} \, d^j_{mq} (\theta),
        \label{eigenfunction}
\end{equation}
which belongs to the eigenvalue
\begin{equation}
        \epsilon_j =  j(j+1) - q^2,
        \qquad
        ( j = |q|, |q|+1, |q|+2, \cdots ).
        \label{eigenvalue}
\end{equation}
Each eigenvalue is degenerated with respect to the index $ m $,
which has a range $ m = -j, -j+1, \cdots, j-1, j $.
Since the lowest eigenvalue of the Laplacian is $ \epsilon_{|q|} = |q| $,
the lower bound of the energy is given by
\begin{equation}
        E 
	\ge
        ( |q| - \mu^2 r^2 )
        \int d \theta d \phi \sin \theta | f_\pm |^2
        + \lambda r^2
        \int d \theta d \phi \sin \theta | f_\pm |^4.
        \label{energy bound}
\end{equation}
If $ |q| - \mu^2 r^2 > 0 $, the RHS of the inequality
(\ref{energy bound}) is positive definite.
Then the minimum of $ E $ is realized only by the trivial vacuum 
$ f \equiv 0 $.
On the other hand,
if $ |q| - \mu^2 r^2 < 0 $, 
it is possible to find a field $ f $ that makes the energy negative.
{}For example, 
let us adopt the eigenfunction (\ref{eigenfunction}) 
of the lowest eigenvalue,
$ f_\pm (\theta,\phi) = c \, e^{-i(m \mp q) \phi} \, d^{|q|}_{mq} (\theta) $,
where $ c $ is a complex number.
The energy of this field configuration
can be written into the form
\begin{equation}
        E 
	= 
        - a_2 |c|^2 + a_4 |c|^4
        \label{negative energy}
\end{equation}
with coefficients $ a_2, a_4 > 0 $.
In particular, we have
$ a_2 = 4 \pi ( 2|q|+1 )^{-1} ( \mu^2 r^2 - |q| ) $.
Thus, for $ 0 < |c|^2 < a_2/a_4 $, this field realizes $ E < 0 $.
Therefore, the minimum value of $ E $ must be negative
and it is realized by a nontrivial vacuum $ f \ne 0 $.
Thus we conclude that 
the rotational symmetry is spontaneously broken 
when the radius $ r $ of $ S^2 $ is larger than the critical radius $ r_q $,
i.e.
\begin{equation}
        r > r_q := \frac{\sqrt{|q|}}{\mu}.
        \label{critical radius}
\end{equation}
We may rephrase this condition 
in terminology of superconductivity physics \cite{vortex}.
In the context of Landau-Ginzburg theory,
the field $ f $ is regarded as the order parameter 
and the energy (\ref{energy}) is regarded as the free energy.
If the field $ f $ develops a nonzero condensation
in the applied magnetic field,
it creates vortices.
The radius of the core of a vortex is characterized
by the coherence length 
$ \xi = M_\sigma^{-1} = (\sqrt{2} \mu)^{-1} $.
Here $ M_\sigma $ is the so-called Higgs mass.
Then
the condition for rotational symmetry breaking (\ref{critical radius})
can be expressed as
\begin{equation}
        r > \sqrt{2|q|} \, \xi.
        \label{coherence length}
\end{equation}
Namely, when the radius of the sphere becomes larger 
than that of the vortex core,
the condensation occurs and the vortices begin to appear.
On the other hand, when the radius is smaller,
the magnetic field is too strong to admit the superconducting state,
and hence the system remains the normal state.

In the present model
the magnetic field is assumed to be a fixed background
and is not allowed to change.
If the reaction to the magnetic field is taken into account,
the Meissner effect should be observed.
We postpone study of the dynamics of the gauge field in the sphere
to future work.

\subsection{Vacuum configurations}
We will now calculate the concrete vacuum configuration of the scalar field.
We will obtain approximate solutions of the scalar field
by a variational method
for three cases with the monopole number $ q = 1/2, 1, 3/2 $.
Let us explain briefly our method of calculation.
A generic field which satisfies the patching condition (\ref{patch})
can be expanded in a series of the eigenfunctions (\ref{eigenfunction}) as
\begin{equation}
        f_{\pm}^{q} (\theta, \phi)
        =
        \sum_{j=|q|}^\infty
        \sum_{m = -j}^j
        c^j_m \, D^{j}_{m,q} (\theta, \phi, \mp \phi).
        \label{expansion}
\end{equation}
Now we use the lowest approximation for it; 
we restrict the series to the leading terms with $ j = |q| $ as
\begin{equation}
        f_{\pm}^{q} (\theta, \phi)
        =
        \sum_{m = -|q|}^{|q|}
        c_m \, D^{|q|}_{m,q} (\theta, \phi, \mp \phi)
        \label{lowest expansion}
\end{equation}
and substitute it into the energy functional (\ref{energy}).
Then the coefficients $ \{ c_m \} $ are adjusted to minimize the energy.
The minimizer is the vacuum configuration.
In the next subsection this lowest approximation will be justified.

\subsubsection{q=1/2}
Let us begin a concrete calculation
for the case of the smallest charge $ q = 1/2 $.
In the expansion (\ref{lowest expansion})
we put 
$ c_{1/2}  = - v e^{ i \gamma /2} \sin (\beta/2) $ and
$ c_{-1/2} =   v e^{-i \gamma /2} \cos (\beta/2) $
with the real parameters $ (v, \beta, \gamma) $.
Then we get
\begin{equation}
        f_{\pm}^{1/2} (\theta, \phi)
        = 
        v \biggl[
                -e^{-i (\phi-\gamma)/2}
                \sin (\beta/2) \cos ( \theta/2 ) 
                +
                e^{i (\phi-\gamma)/2} 
                \cos (\beta/2) \sin ( \theta/2 ) 
        \biggr]
        e^{\pm i \phi/2}.
        \label{ground state for q=1/2}
\end{equation}
Thus it can easily be seen that
the value of $ f^{1/2} $ vanishes
at the point $ (\theta, \phi) = (\beta, \gamma) $.
Since the position of the zero point of $ f^{1/2} $ can be moved
to the north pole of $ S^2 $ by an appropriate $ SU(2) $ rotation,
we can set $(\beta, \gamma) = (0, 0)$ without loss of generality.
Then the vacuum field is given by
\begin{equation}
        \bra f_{\pm}^{1/2} (\theta, \phi) \ket
        = 
        v \, e^{i (1 \pm 1) \phi /2} \sin ( \theta/2 ),
        \label{ground for 1/2}
\end{equation}
\begin{figure}[tbhp]
\includegraphics{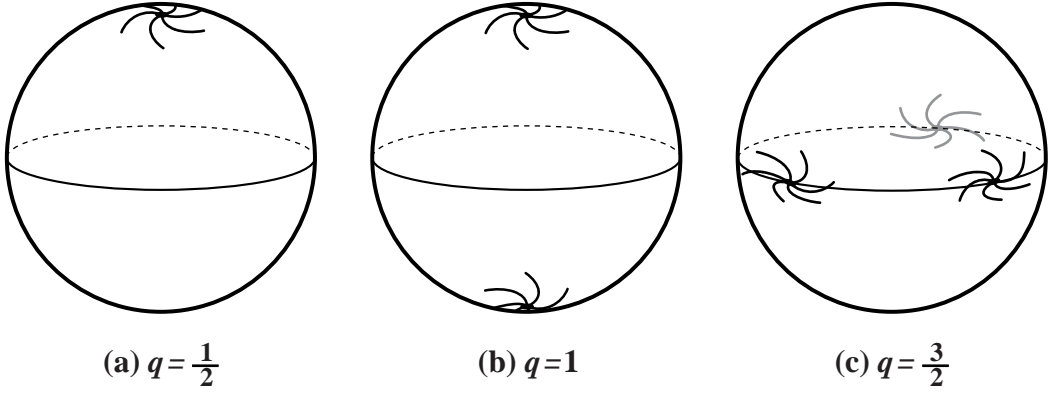}
\caption{Vortex configurations}
\end{figure}
and the single vortex is located at the north pole of $ S^2 $.
The configurations of vortices for various $ q $ are shown in Fig.1.
The energy (\ref{energy}) is then calculated as
\begin{equation}
        E 
        =
        - 2 \pi \left( \mu^2 r^2 - \frac{1}{2} \right) v^2
        + 
        \frac{2}{3} \cdot 2 \pi \lambda r^2 v^4.
        \label{variational energy for q=1/2}
\end{equation}
Hence the energy is minimized by
\begin{equation}
	v^2 = 
	\left\{
		\begin{array}{ll}
		    0 
		    & \quad \mbox{for} \quad r \leq \sqrt{1/2} \, \mu^{-1}, 
		    \\
		    \Big( 1 - \dfrac{1}{2 \mu^2 r^2 } \Big) 
		    \dfrac{3 \mu^2}{4 \lambda} 
		    & \quad \mbox{for} \quad r > \sqrt{1/2} \, \mu^{-1}.
		\end{array} 
	\right.
        \label{VEV for q=1/2}
\end{equation}
This result is to be compared with the vacuum expectation value
\begin{equation}
        \bra f^0 \ket^2 
        = \frac{\mu^2}{2 \lambda} 
        \label{VEV for q=0}
\end{equation}
for the case of $ q=0 $.

It can be verified that
the vacuum field configuration (\ref{ground for 1/2}) is invariant under
the combined transformation 
of 
the rotation (\ref{f to f'}) of the angle $ \gamma $ around the $ z $-axis
with 
the phase rotation (\ref{U(1)}) $ t = - \gamma $
\begin{equation}
	e^{i \gamma/2}
	\varphi (e^{- i \sigma_3 \gamma/2}) :
        f_\pm^{1/2} (\theta,\phi)
        \; \mapsto \;
        e^{    i \gamma/2} 
        e^{\pm i \gamma/2} f_\pm^{1/2} (\theta,\phi - \gamma).
        \label{remaining U(1)}
\end{equation}
This transformation
$ e^{i \gamma/2} \varphi (e^{- i \sigma_3 \gamma/2}) $
is equal to $ e^{-i (L_3 - Q) \gamma} $.
On the other hand,
it is obvious that 
$ \bra f^{1/2} \ket $ is not invariant under $ CP $.
Therefore, we conclude that 
when $ q = 1/2 $ and $ r > ( \sqrt{2} \mu )^{-1} $,
the symmetry $ U(1) \times SU(2) \times CP $ 
is spontaneously broken to the subgroup 
$ U(1)' $ that is generated by $ L_3 - Q $.
Accordingly there are 
three massless 
Nambu-Goldstone bosons,
which couple to the generators 
$ \{ X_1, X_2, X_3 \} = \{ L_1, L_2, L_3 + Q \} $ of the broken symmetry.
However, the Nambu-Goldstone boson 
coupling to the charge $ Q $
would be absorbed into the gauge boson via the Higgs mechanism 
if the gauge field has its own dynamical degrees of freedom.
The scalar field $ f(x^0, x^1, \cdots, x^{n-1}, \theta, \phi) $
is expanded around the vacuum (\ref{ground for 1/2}) as
\begin{eqnarray}
	f(x^0, x^1, \cdots, x^{n-1}, \theta, \phi )
	& = &
	\bra f(\theta, \phi) \ket
	-i
	\sum_j  X_j \bra f(\theta, \phi) \ket \,
	\pi^j ( x^0, x^1, \cdots, x^{n-1} )
	\nonumber \\ &&
	+
	f_{\mbox{\tiny massive}} (x^0, x^1, \cdots, x^{n-1}, \theta, \phi ).
	\label{Nambu-Goldstone 1/2}
\end{eqnarray}
The three real fields $ \{ \pi^1, \pi^2, \pi^3 \} $ describe
the Nambu-Goldstone bosons.
We can calculate each mode by operating each generator,
(\ref{L_3}), (\ref{L+}), (\ref{L-}),
on the vacuum (\ref{ground for 1/2}) as
\begin{eqnarray}
	L_+ \bra f^{1/2} _\pm \ket
& = &	0,
	\label{L+ f^1/2} 
\\
	L_- \bra f^{1/2} _\pm \ket
& = &	- v e^{i (-1 \pm 1) \phi /2} \cos ( \theta/2 ),
	\label{L- f^1/2} 
\\
	(L_3 + Q) \bra f^{1/2} _\pm \ket
& = &	v \, e^{i (1 \pm 1) \phi /2} \sin ( \theta/2 ),
	\label{L3+Q f^1/2} 
\\
	(L_3 - Q) \bra f^{1/2} _\pm \ket
& = &	0.
	\label{L3-Q f^1/2} 
\end{eqnarray}
Eq.(\ref{L3-Q f^1/2}) reflects the invariance
$ e^{-i (L_3 - Q) \gamma} \bra f_\pm^{1/2} \ket = \bra f_\pm^{1/2} \ket $.
Using Eq.(\ref{L+ f^1/2}) 
we rewrite the expansion (\ref{Nambu-Goldstone 1/2}) as
\begin{eqnarray}
	f^{1/2}_\pm (x, \theta, \phi)
& = &
	\bra f^{1/2}_\pm (\theta, \phi) \ket
	- \frac{i}{2} L_- \bra f^{1/2} _\pm \ket \,
	( \pi^1 + i \pi^2 ) ( x )
	- i (L_3 + Q) \bra f^{1/2} _\pm \ket \,
	\pi^3 ( x )
	\nonumber \\ &&
	+
	( f^{1/2}_{\mbox{\tiny massive}} )_\pm
	(x,\theta, \phi).
	\label{Nambu-Goldstone 1/2'}
\end{eqnarray}
Besides, the massive modes are further expanded as
\begin{eqnarray}
	( f^{1/2}_{\mbox{\tiny massive}} )_\pm 
	(x,\theta, \phi)
& = &
	\frac{1}{2} L_- \bra f^{1/2} _\pm \ket \,
	( \sigma^1 + i \sigma^2 ) ( x )
	+ (L_3 + Q) \bra f^{1/2} _\pm \ket \,
	\sigma^3 ( x )
	\nonumber \\ &&
	+
	\sum_{j = 3/2}^{\infty} \sum_{m = -j}^j
	D^j_{m,1/2} (\theta, \phi, \mp \phi) S_j^m (x)
	\label{massive mode 1/2}
\end{eqnarray}
with real scalars $ \sigma^j (x) $ and complex scalars $ S_j^m(x) $.

We would like to put a remark about a relation of our argument with
the Coleman theorem \cite{Coleman},
which forbids spontaneous breaking of continuous symmetries
in two dimensions.
Our model is built on higher dimensions than two;
it is in the direct product space $ \R^n \times S^2 $
of the Minkowski space $ \R^n $ with the extra $ S^2 $.
Thus the Coleman theorem is not applicable to our model.

\subsubsection{q=1}
Next let us consider the case of $ q=1 $.
Then a generic scalar field $ f $ has two vortices on $ S^2 $.
The lowest expansion (\ref{lowest expansion}) now becomes
\begin{eqnarray}
        f_{\pm}^1 (\theta, \phi)
& = &
        \frac{1}{2}
        \left[
                c_{+1} \, e^{-i \phi} ( 1 + \cos \theta ) 
                +
                c_{0}  \, \sqrt{2} \sin \theta
                +
                c_{-1} \, e^{i \phi} ( 1 - \cos \theta ) 
        \right] e^{\pm i \phi}
	\nonumber \\
& = &
	\frac{1}{\sqrt{2}}
	\Bigl[
		c_x 
		( i \sin \phi - \cos \theta \, \cos \phi )
		+
		c_y
		( -i \cos \phi - \cos \theta \, \sin \phi )
	\nonumber \\ && \qquad
		+
		c_z \sin \theta
	\Bigr] e^{\pm i \phi}.
	\label{ground state for q=1}
\end{eqnarray}
We have put 
$ c_{\pm 1} = \mp (c_x \pm ic_y) / \sqrt{2} $ 
and $ c_0 = c_z $.
Under the $ SU(2) $ transformation (\ref{SU(2)f}),
the vector $ (c_x,c_y,c_z) $ obeys the triplet representation.
By using the $ U(1) \times SU(2) $ transformations, 
(\ref{U(1)}) and (\ref{SU(2)f}),
we can turn the coefficients into the form
\begin{equation}
	( c_x, c_y, c_z )
	=
	( 
		0,
		iv \sin \alpha,
		v  \cos \alpha
	),
	\label{c vector canonical form}
\end{equation}
with the real parameters $ v $ and $ \alpha\ (0\leq \alpha \leq \pi/4) $.
Then the field (\ref{ground state for q=1}) becomes
\begin{equation}
        f_{\pm}^1 (\theta, \phi)
	= 
        \frac{1}{\sqrt{2}} \, v
        \Bigl[
                  ( \cos \alpha \sin \theta + \sin \alpha \cos \phi )
                - i \sin \alpha \cos \theta \sin \phi 
        \Bigr] e^{\pm i \phi}.
        \label{ground state for q=1 by alpha}
\end{equation}
Thus two zero points of $f^1$ are located at
$ (\theta, \phi) = ( \sin^{-1}(\tan\alpha), \pi) $.
So, the relative displacement of two vortices is controlled 
by the parameter $ \alpha $.

Substituting the trial function (\ref{ground state for q=1 by alpha}) 
into (\ref{energy}),
we evaluate the total energy as
\begin{equation}
        E 
        =
        - \frac{4 \pi}{3} ( \mu^2 r^2 - 1 ) v^2
        + \frac{8 \pi}{15} 
        \left\{
                1 + \frac{1}{4} ( 1 - \cos 4 \alpha )
        \right\}
        \lambda r^2 v^4.
        \label{total energy of f_1}
\end{equation}
The minimum of the potential is realized when
$ \cos 4 \alpha = 1 $, i.e. 
$ \alpha = 0 $.
Then the two vortices are located at opposite two points on $ S^2 $.
On the other hand, the maximum of the potential is realized when
$ \cos 4 \alpha = -1 $, i.e. 
$ \alpha = \pi/4$.
Then the two vortices coincide.
Thus we observe that {\it the vortices repel each other.}
The minimum of the energy is realized by $ \alpha = 0 $ with
\begin{equation}
        v^2 =
	\left\{
		\begin{array}{ll}
		0 
		& \quad \mbox{for} \quad r \leq \mu^{-1}, 
		\\
		\Big( 1 - \dfrac{1}{\mu^2 r^2} \Big) 
		\dfrac{5 \mu^2}{4 \lambda} 
        	& \quad \mbox{for} \quad r > \mu^{-1}.
        	\end{array}
        \right.
        \label{VEV of q=1}
\end{equation}
The vacuum configuration (\ref{ground state for q=1 by alpha}) 
then becomes
\begin{equation}
        \bra f_{\pm}^1 ( \theta, \phi) \ket
        =
        \frac{1}{\sqrt{2}} \, v \sin \theta \, e^{\pm i \phi},
        \label{opposite vortex}
\end{equation}
and has two vortices at the north and the south poles of $ S^2 $, respectively.
This configuration (\ref{opposite vortex})
is invariant under the rotations around the $ z $-axis (\ref{f to f'}),
which is generated by $ L_3 $.
Under $ CP $ given in (\ref{CP f}), $ \bra f^1 \ket $ is transformed as
\begin{equation}
	( \varphi_{CP}  \bra f^1 \ket )_{\pm} ( \theta, \phi)
	=
	- \frac{1}{\sqrt{2}} \, v \, e^{\pm i (\pi + \phi)} \sin ( \pi-\theta )
	=
	\frac{1}{\sqrt{2}} \, v \, e^{\pm i \phi} \sin \theta.
	\label{invariance of 1}
\end{equation}
Therefore, $ \bra f^1 \ket $ is $ CP $ invariant.
Thus, we conclude that when $ q = 1 $ and $ r > \mu^{-1} $,
two vortices appear and settle down at two diametrical points on $ S^2 $.
The symmetry $ U(1) \times SU(2) \times CP $ is spontaneously broken to 
$ U(1)'' \times CP $, where $ U(1)''$ is generated by $ L_3 $.
In the broken phase
three Nambu-Goldstone bosons,
which couple to the generators $ \{ Q, L_1, L_2 \} $,
appear.
The field $ f $ can be expanded in a way similar to (\ref{Nambu-Goldstone 1/2})
as 
\begin{eqnarray}
	f^{1}_\pm (x,\theta,\phi)
& = &
	\bra f^{1}_\pm (\theta, \phi) \ket
	- \frac{i}{2} L_+ \bra f^{1} _\pm \ket \,
	( \pi^1 - i \pi^2 ) ( x )
	- \frac{i}{2} L_- \bra f^{1} _\pm \ket \,
	( \pi^1 + i \pi^2 ) ( x )
	\nonumber \\ &&
	- i Q \bra f^{1} _\pm \ket \,
	\pi^0 ( x )
	+
	( f^{1}_{\mbox{\tiny massive}} )_\pm
	(x,\theta, \phi),
	\label{Nambu-Goldstone 1}
\end{eqnarray}
where $ \{ \pi^0, \pi^1, \pi^2 \} $ are
the Nambu-Goldstone fields.
Each mode is calculated as
\begin{eqnarray}
	Q \bra f^1_\pm \ket
& = &
	\frac{1}{\sqrt{2}} \, v \, e^{\pm i \phi} \sin \theta,
	\label{Q f1}
\\
	L_+ \bra f^1_\pm \ket
& = &
	- \frac{1}{\sqrt{2}} \, v e^{i (1 \pm 1) \phi} ( 1 - \cos \theta ),
	\label{L+ f1} 
\\
	L_- \bra f^1_\pm \ket
& = &
	- \frac{1}{\sqrt{2}} \, v e^{ i ( -1 \pm 1 ) \phi } ( 1 + \cos \theta ),
	\label{L- f1}
\\
	L_3 \bra f^1_\pm \ket
& = &	0.
	\label{L3 f1}
\end{eqnarray}
The last equation (\ref{L3 f1}) confirms
the invariance $ e^{-i L_3 \gamma} \bra f^1_\pm \ket = \bra f^1_\pm \ket $.
However, the boson of $ \pi^0 $, which couples to the charge $ Q $,
would disappear via the Higgs mechanism,
if the gauge field has its own dynamical degrees of freedom.
Furthermore, the massive modes are given by
\begin{eqnarray}
	( f^{1}_{\mbox{\tiny massive}} )_\pm 
	(x,\theta, \phi)
& = &
	\frac{1}{2}   L_+ \bra f^{1} _\pm \ket \, ( \sigma^1 - i \sigma^2 ) (x)
	+ \frac{1}{2} L_- \bra f^{1} _\pm \ket \, ( \sigma^1 + i \sigma^2 ) (x)
	\nonumber \\ &&
	+ Q \bra f^{1} _\pm \ket \, \sigma^0 ( x )
	+ 
	\sum_{j = 2}^{\infty} \sum_{m = -j}^j
	D^j_{m,1} (\theta, \phi, \mp \phi) S_j^m (x)
	\label{massive mode 1}
\end{eqnarray}
with real scalars $ \sigma^j (x) $ and complex scalars $ S_j^m(x) $.

\subsubsection{q=3/2}
{}Finally, let us examine the case of $ q=3/2 $.
Then the number of vortices is three.
The lowest expansion (\ref{lowest expansion}) now becomes
\begin{eqnarray}
        f_{\pm}^{3/2} (\theta, \phi)
& = &
	\Big[
	  c_{3/2}           e^{-3i \phi /2 } \cos^3 (\theta/2)
	+ c_{1/2}  \sqrt{3} e^{- i \phi /2 } \cos^2 (\theta/2) \sin (\theta/2)
	\nonumber \\ && 
	+ c_{-1/2} \sqrt{3} e^{  i \phi /2 } \cos (\theta/2) \sin^2 (\theta/2)
	+ c_{-3/2}          e^{ 3i \phi /2 } \sin^3 (\theta/2)
	\Big] e^{\pm 3i \phi/2}.
	\qquad
	\label{expansion 3/2}
\end{eqnarray}
By substituting it into (\ref{energy}) the energy functional is evaluated as
\begin{eqnarray}
	E
& = &
	- \pi \bigg( \mu^2 r^2 - \dfrac{3}{2} \bigg)
	\Big(
		|c_{ 3/2}|^2 +
		|c_{ 1/2}|^2 +
		|c_{-1/2}|^2 +
		|c_{-3/2}|^2
	\Big)
	\nonumber \\ &&
	+ 
	\frac{4}{35} \pi r^2 \lambda
	\Big\{
		5 |c_{ 3/2}|^4 +
		3 |c_{ 1/2}|^4 +
		3 |c_{-1/2}|^4 +
		5 |c_{-3/2}|^4 
	\nonumber \\ && \quad
		+ 10 |c_{ 3/2}|^2 |c_{ 1/2}|^2 
		+ 4  |c_{ 3/2}|^2 |c_{-1/2}|^2 
		+    |c_{ 3/2}|^2 |c_{-3/2}|^2
	\nonumber \\ && \quad
		+ 9  |c_{ 1/2}|^2 |c_{-1/2}|^2
		+ 4  |c_{ 1/2}|^2 |c_{-3/2}|^2
		+ 10 |c_{-1/2}|^2 |c_{-3/2}|^2 
	\nonumber \\ && \quad
		+ 3 c_{ 1/2} c_{-1/2} c_{ 3/2}^* c_{-3/2}^* 
		+ 3 c_{ 1/2}^* c_{-1/2}^* c_{ 3/2} c_{-3/2} 
	\nonumber \\ && \quad
		+ 2 \sqrt{3} \, c_{-1/2} c_{3/2} (c_{1/2}^* )^2
		+ 2 \sqrt{3} \, c_{-1/2}^* c_{3/2}^* (c_{1/2})^2
	\nonumber \\ && \quad
		+ 2 \sqrt{3} \, c_{ 1/2}^* c_{-3/2}^* (c_{-1/2} )^2
		+ 2 \sqrt{3} \, c_{ 1/2} c_{-3/2} (c_{-1/2}^*)^2
	\Big\}.
	\label{E 3/2}
\end{eqnarray}
The minimum of $ E $ is then found at
$ ( c_{3/2}, c_{1/2}, c_{-1/2}, c_{-3/2} ) = (-v, 0, 0, v ) $,
which is unique up to the $ U(1) \times SU(2) $ symmetry.
Thus the vacuum is
\begin{eqnarray}
        \bra f_{\pm}^{3/2} (\theta,\phi) \ket
& = &
        v 
        \Big[
		- e^{-3i \phi /2} \cos^3 ( \theta/2 ) 
		+ e^{ 3i \phi /2} \sin^3 ( \theta/2 ) 
        \Big]
        e^{\pm 3 i \phi /2 }
        \nonumber \\
& = &
        v 
        \Big[
                  e^{-i \phi /2} \cos ( \theta/2 ) 
                - e^{ i \phi /2} \sin ( \theta/2 ) 
        \Big]
        \nonumber \\ && 
        \times \Big[
                  e^{-i (\phi - 2 \pi/3) /2} \cos ( \theta/2 ) 
                - e^{ i (\phi - 2 \pi/3) /2} \sin ( \theta/2 ) 
        \Big]
        \nonumber \\ && 
        \times \Big[
                  e^{-i (\phi - 4 \pi/3) /2} \cos ( \theta/2 ) 
                - e^{ i (\phi - 4 \pi/3) /2} \sin ( \theta/2 ) 
        \Big]
        e^{\pm 3 i \phi /2 } 
        \label{vacuum of q=3/2}
\end{eqnarray}
with
\begin{equation}
	v^2 = 
	\left\{
		\begin{array}{ll}
		0 
		& \quad \mbox{for} \quad r \leq \sqrt{3/2} \, \mu^{-1}, 
		\\
		\Big( 1 - \dfrac{3}{2 \mu^2 r^2} \Big) 
		\dfrac{35 \mu^2}{44 \lambda}
		& \quad \mbox{for} \quad r > \sqrt{3/2} \, \mu^{-1}.
		\end{array} 
	\right.
        \label{v of q=3/2}
\end{equation}
We can read off the location of zero points from (\ref{vacuum of q=3/2});
they are located at $ \phi = 0, 2\pi/3, 4\pi/3 $
on $ \theta = \pi/2 $.
Namely, the vortices are located at the vertices
of the largest equilateral triangle on the equator
as shown in Fig.1(c).
Then the vacuum energy is evaluated as
\begin{eqnarray}
	E_{\mbox{\tiny{vac}}}
& = &
	- 
	\bigg( \mu^2 r^2 - \frac{3}{2} \bigg)^2
	\frac{35 \pi}{44 \lambda r^2}.
\end{eqnarray}

For comparison
let us calculate the energy of the configuration 
in that the three vortices coincide at the north pole of $ S^2 $.
Such a configuration is specified by
$ ( c_{3/2}, c_{1/2}, $ $ c_{-1/2}, c_{-3/2} ) = (0, 0, 0, v' ) $.
Then the energy (\ref{E 3/2}) is 
\begin{eqnarray}
	E =
	- \pi \bigg( \mu^2 r^2 - \dfrac{3}{2} \bigg) v'^2
	+ 
	\frac{20}{35} \pi r^2 \lambda v'^4.
	\label{E 3/2 coinciding vortex}
\end{eqnarray}
The minimal of $ E $ is realized by
\begin{equation}
	v'^2 = 
	\bigg( 1 - \dfrac{3}{2 \mu^2 r^2} \bigg) 
	\dfrac{7 \mu^2}{8 \lambda}
	\quad \mbox{for} \quad r > \sqrt{3/2} \, \mu^{-1}
        \label{v' of q=3/2}
\end{equation}
and the minimal value is
\begin{equation}
	E_{\mbox{\tiny{coincide}}}
	=
	- 
	\frac{11}{20} \cdot
	\bigg( \mu^2 r^2 - \frac{3}{2} \bigg)^2
	\frac{35 \pi}{44 \lambda r^2}.
\end{equation}
Hence we can see that
$ E_{\mbox{\tiny{coincide}}} > E_{\mbox{\tiny{vac}}} $.
Therefore 
{\it the vortex whose topological number is three is unstable
and decays into three separated vortices.}
This result corresponds to Type II superconductor, 
in which vortices repel each other and form the Abrikosov lattice \cite{vortex}.
Jacobs and Rebbi \cite{Jacobs} gave a detailed analysis
of interaction of vortices of the Abelian Higgs model in $ \R^2 $.

Now we describe symmetry of the vacuum.
The rotation around the $ z $-axis (\ref{f to f'})
of the angle $ \gamma = 2 \pi/3 $
followed by
the $ U(1) $ rotation (\ref{U(1)}) 
of the phase $ t = 2 \pi/3 $,
transforms the scalar field as
\begin{equation}
        \varphi_R 
	:= e^{-i \pi} \varphi (e^{-i \sigma_3 \pi/3}) :
        f_\pm^{3/2} (\theta,\phi)
        \; \mapsto \;
        f_\pm^{3/2} (\theta,\phi - 2 \pi/3).
        \label{R}
\end{equation}
This composite transformation $ \varphi_R $ generates the cyclic group $ \Z_3 $.
Actually, the configuration (\ref{vacuum of q=3/2}) remains invariant under
the operation of $ \varphi_R $.
On the other hand,
the $ \pi $-rotation around the $ x $-axis (\ref{f to f' x-axis})
followed by
the $ U(1) $ rotation (\ref{U(1)}) 
of the phase $ t = \pi/3 $
transforms the scalar field as
\begin{equation}
        \varphi_T 
	:= e^{-i \pi/2} \varphi (e^{-i \sigma_1 \pi/2}) :
        f_\pm^{3/2} (\theta,\phi)
        \; \mapsto \;
        e^{i \pi} 
        f_\mp^{3/2} (\pi-\theta, -\phi).
        \label{T}
\end{equation}
This composite transformation $ \varphi_T $ generates 
another cyclic group $ \Z_2 $.
It can easily be verified that the configuration
(\ref{vacuum of q=3/2}) remains invariant also under the operation of 
$ \varphi_T $.
Note that the two operations $ \varphi_R $ and $ \varphi_T $ 
do not commute each other;
they generate a nonabelian group $ D_3 $, 
which is called the dihedral group of the order three,
i.e. the symmetry group of a regular triangle.

Moreover, the vacuum for $ q = 3/2 $ 
has another discrete symmetry that originates from $ CP $.
The vacuum configuration $ \bra f^{3/2} \ket $ is transformed by $ CP $ as
\begin{equation}
        ( \varphi_{CP} \bra f^{3/2} \ket )_{\pm} (\theta,\phi)
	=
        i v 
        \Big[
		  e^{-3i \phi /2} \cos^3 ( \theta/2 ) 
		+ e^{ 3i \phi /2} \sin^3 ( \theta/2 ) 
        \Big]
        e^{\pm 3 i \phi /2 }.
        \label{CP 3/2}
\end{equation}
Hence $ \bra f^{3/2} \ket $ is not invariant under $ CP $.
To turn it into the original form we apply 
the rotation around the $ z $-axis (\ref{f to f'}) with $ \gamma = \pi $
and the phase rotation (\ref{U(1)}) with $ t = 2 \pi/3 $ to get
\begin{equation}
	e^{-i \pi} \varphi (e^{-i \sigma_3 \pi/2}) :
	f^{3/2}_\pm (\theta,\phi) 
	\; \mapsto \;
	- e^{\pm 3 i \pi/2} f^{3/2}_\pm (\theta, \phi - \pi).
	\label{z-axis pi}
\end{equation}
Then we can verify the invariance
\begin{equation}
        ( e^{-i \pi} \varphi (e^{-i \sigma_3 \pi/2}) \circ \varphi_{CP} 
        \bra f^{3/2} \ket )_{\pm} 
        (\theta,\phi)
	=
	\bra f^{3/2}_{\pm} \ket (\theta,\phi).
        \label{CP + z-rot}
\end{equation}
Note that the parity followed by
the $ \pi $-rotation around the $ z $-axis is equivalent to
the reflection by a mirror perpendicular to the $ z $-axis:
\begin{equation}
	\left(
		\begin{array}{l} x \\ y \\ z \end{array}
	\right)
	\mapright{P}{}
	\left(
		\begin{array}{l} -x \\ -y \\ -z \end{array}
	\right)
	\mapright{e^{-i \sigma_3 \pi/2}}{}
	\left(
		\begin{array}{r} x \\ y \\ -z \end{array}
	\right).
	\label{reflection}
\end{equation}
Thus the symmetry of the vacuum of $ q=3/2 $ includes
the reflection symmetry.

We thus conclude that
when $ q = 3/2 $ and $ r > \sqrt{3/2} \, \mu^{-1} $,
three vortices appear and settle down at three points
separated furthest each other.
The symmetry $ U(1) \times SU(2) \times CP $ is spontaneously broken to 
the discrete nonabelian group $ D_{3h} $.
Here $ D_{3h} $ denotes 
the dihedral group of the order three with the horizontal reflection.
In plain words, $ D_{3h} $ is the symmetry group of a regular triangle
without orientation.
Hence
there are four Nambu-Goldstone bosons,
which couple to the generators $ \{ Q, L_1, L_2, L_3 \} $ 
of the broken symmetry.
The field $ f $ can be expanded in a way similar to (\ref{Nambu-Goldstone 1/2})
as 
\begin{eqnarray}
	f^{3/2}_\pm (x,\theta, \phi)
& = &
	\bra f^{3/2}_\pm (\theta, \phi) \ket
	- \frac{i}{2} L_+ \bra f^{3/2} _\pm \ket \, ( \pi^1 - i \pi^2 ) ( x )
	- \frac{i}{2} L_- \bra f^{3/2} _\pm \ket \, ( \pi^1 + i \pi^2 ) ( x )
	\nonumber \\ &&
	- \frac{i}{2} (L_3 - Q) \bra f^{3/2} _\pm \ket \, (\pi^3 - \pi^0) (x)
	- \frac{i}{2} (L_3 + Q) \bra f^{3/2} _\pm \ket \, (\pi^3 + \pi^0) (x)
	\nonumber \\ &&
	+
	( f^{3/2}_{\mbox{\tiny massive}} )_\pm
	(x,\theta, \phi),
	\label{Nambu-Goldstone 3/2}
\end{eqnarray}
where $ \{ \pi^0, \pi^1, \pi^2, \pi^3 \} $ are
massless fields associated with the modes
\begin{eqnarray}
	( L_3 - Q ) \bra f^{3/2}_\pm \ket
& = &
        3 v \, e^{-3 i \phi /2} e^{\pm 3 i \phi /2 } \cos^3 ( \theta/2 ),
	\label{L3-Q f3/2}
\\
	L_+ \bra f^{3/2}_\pm \ket
& = &
	3 v \, e^{-i \phi /2} e^{\pm 3 i \phi /2 }
	\cos^2 ( \theta/2 ) \sin ( \theta/2 ),
	\label{L+ f3/2} 
\\
	L_- \bra f^{3/2}_\pm \ket
& = &
	-3 v \, e^{i \phi /2} e^{\pm 3 i \phi /2 } 
	\cos ( \theta/2 ) \sin^2 ( \theta/2 ),
	\label{L- f3/2}
\\
	( L_3 + Q ) \bra f^{3/2}_\pm \ket
& = &
        3 v \, e^{ 3i \phi /2} e^{\pm 3 i \phi /2 } \sin^3 ( \theta/2 ).
	\label{L3+Q f3/2}
\end{eqnarray}
Actually, the boson of $ \pi^0 $, which couples to $ Q $,
would be absorbed into the gauge boson via the Higgs mechanism.

Our results are summarized as follows:
let $ G $ be the original symmetry of the model
and $ H_q $ be the symmetry of the vacuum for the monopole number $ q $.
Then we have observed the pattern of symmetry breaking
\begin{equation}
	G = U(1) \times SU(2) \times CP
	\to
	\left\{
		\begin{array}{ll}
		H_{1/2} = U(1)' 
		& \mbox{for} \quad r > \sqrt{1/2} \, \mu^{-1} \\
		H_{1}   = U(1)'' \times CP     \quad
		& \mbox{for} \quad r > \mu^{-1} \\
		H_{3/2} = D_{3h}
		& \mbox{for} \quad r > \sqrt{3/2} \, \mu^{-1} \\
		\end{array}
	\right.
\end{equation}
where 
$ U(1)' = \{ e^{-i (L_3 - Q) \gamma} \} $ and
$ U(1)''= \{ e^{-i L_3 \gamma} \} $.

Before closing this subsection, 
we should make a comment on $ H_q $ for $ q > 3/2 $. 
Although it is difficult 
to continue to find vacuum configurations for $ q > 3/2 $
even in the lowest approximation (\ref{lowest expansion}),
it seems reasonable to speculate that 
the unbroken symmetry $ H_q $ is discrete for $ q \ge 3/2 $. 
This is expected from the observations that 
$ 2q $ vortices appear for $ r > r_q $ and 
that the potential between vortices is repulsive. 
Then, three separated vortices are enough 
to break the continuous symmetry $ G $ to a discrete symmetry.

\subsection{Stability of the vacuum}
In the previous subsection
we solved the problem to minimize the energy functional (\ref{energy})
by the variational method within the restricted function space
(\ref{lowest expansion}),
which is the eigenspace 
belonging to the lowest eigenvalue of the monopole Laplacian (\ref{Laplacian}).
Here we would like to clarify validity of our analysis.

The first point to be declared is 
that the critical radius (\ref{critical radius}) is exact
in the context of the classical field theory.
We did not recourse any approximation 
in the evaluation of the critical radius.

The second point to be examined is 
accuracy of the approximate solutions of the vacuum, like
(\ref{ground for 1/2}),
(\ref{opposite vortex}),
(\ref{vacuum of q=3/2}).
We calculated them by the variational method
with restricting
the full function space (\ref{expansion})
to the lowest eigenspace (\ref{lowest expansion})
of the monopole Laplacian.
This restriction is a good approximation
if the functions belonging to the lowest eigenvalue $ \epsilon_{|q|} $ 
give
negative contribution to the quadratic term in (\ref{energy}),
whereas
those belonging to the next eigenvalue $ \epsilon_{|q|+1} $ 
give positive contribution to it.
More explicitly, 
referring to the eigenvalues (\ref{eigenvalue}),
the necessary condition for the validity is
\begin{equation}
        \epsilon_{|q|} - \mu^2 r^2 
        < 0 <
        \epsilon_{|q|+1} - \mu^2 r^2, 
        \quad \mbox{or} \quad
        \frac{\sqrt{|q|}}{\mu} 
        < r < 
        \frac{\sqrt{3|q| + 2}}{\mu}.
        \label{valid range}
\end{equation}
In other words,
the restricted variational method 
is valid
when the radius of $ S^2 $ is larger than the critical radius 
but not too large.

The third point to be questioned is the stability of the vacuum 
against perturbations by higher eigenvalue functions.
The approximation is improved by including higher-order terms in the expansion
(\ref{expansion}).
Now a question arises:
if we include some of or all of higher terms in the expansion
(\ref{expansion}),
does the better-approximated or the precise vacuum have the same symmetry 
as the lowest-approximated vacuum has?

The answer is affirmative:
when higher-order terms are included in the trial function
(\ref{expansion}),
but if the radius of $ S^2 $ is in the range (\ref{valid range}),
the vacuum calculated by the higher expansion has the same symmetry
as the vacuum calculated by the lowest approximation has.

The above statement is proved as follows:
{}first,
note that the space of the scalar fields $ f $
provides a unitary representation of $ U(1) \times SU(2) $
by obeying the transformations (\ref{U(1)}) and (\ref{SU(2)f}).
Let $ f^{(0)} $ be the solution obtained by the lowest approximation like
(\ref{ground for 1/2}),
(\ref{opposite vortex}),
(\ref{vacuum of q=3/2}).
Assume that $ f $ is a solution obtained by the higher-order approximation.
Then define $ f^{(1)} $ as a correction to $ f^{(0)} $, i.e.
\begin{equation}
        f = f^{(0)} + f^{(1)}.
        \label{difference}
\end{equation}
Let $ H $ be the subgroup of $ U(1) \times SU(2) $
that preserves $ f^{(0)} $ invariant.
Then we decompose
$ f^{(1)} $
into a component
$ ( f^{(1)} )_\parallel $ that is in the identity representation of $ H $,
and its orthogonal complement
$ ( f^{(1)} )_\perp $
as
\begin{equation}
        f^{(1)} =
        ( f^{(1)} )_\parallel + ( f^{(1)} )_\perp.
        \label{decompose}
\end{equation}
When (\ref{difference}) is substituted,
the energy functional (\ref{energy}) is symbolically written as
\begin{equation}
        E [ f^{(0)} + f^{(1)} ] 
        =
        E [ f^{(0)} ] 
        + \frac{\delta E}{\delta f} [ f^{(0)} ] \cdot f^{(1)}
        + \frac{1}{2} \frac{\delta^2 E}{\delta f^2} [ f^{(0)} ] 
        \cdot (f^{(1)})^2
        + \cdots,
        \label{expand E}
\end{equation}
which is regarded as a polynomial with respect to $ f^{(1)} $.
The second term of the RHS is
\begin{eqnarray}
        \frac{\delta E}{\delta f} [ f^{(0)} ] \cdot f^{(1)}
        & = &
        \int d \theta d \phi \sin \theta
        \Big\{
                f^{(0)*} 
                ( - \Delta_q - \mu^2 r^2 
                + 2 \lambda r^2 | f^{(0)} |^2 ) 
                f^{(1)}
        \nonumber \\ && \qquad \qquad \quad
                +
                f^{(1)*} 
                ( - \Delta_q - \mu^2 r^2 
                + 2 \lambda r^2 | f^{(0)} |^2 ) 
                f^{(0)}
        \Big\}.
        \label{1st order}
\end{eqnarray}
Since $ f^{(0)} $ is invariant under the actions of $ H $,
$ ( - \Delta_q - \mu^2 r^2 + 2 \lambda r^2 |f^{(0)}|^2 ) f^{(0)} $ is also
invariant.
Therefore, the term linear in the orthogonal component $ (f^{(1)})_\perp $
vanishes as
\begin{equation}
        \frac{\delta E}{\delta f} [ f^{(0)} ] \cdot (f^{(1)})_\perp
        = 0.
        \label{orth}
\end{equation}
Moreover, the third term of the RHS of (\ref{expand E}) is
\begin{eqnarray}
        \frac{1}{2} \frac{\delta^2 E}{\delta f^2} [ f^{(0)} ] 
        \cdot ( f^{(1)} )^2
& = &
        \int d \theta d \phi \sin \theta 
        \bigg[
                f^{(1)*} 
                ( - \Delta_q - \mu^2 r^2 ) 
                f^{(1)}
                \nonumber \\ &&
                + \lambda r^2
                \Big\{
                          2 | f^{(0)} |^2 | f^{(1)} |^2
                        + \big( f^{(0)}  f^{(1)*} + f^{(0)*} f^{(1)} \big)^2
                \Big\}
        \bigg].
        \label{2nd order}
\end{eqnarray}
If the radius of $ S^2 $ is in the range (\ref{valid range}),
the first term 
$ \int f^{(1)*} ( - \Delta_q - \mu^2 r^2 ) f^{(1)} $
in the RHS
is positive definite for $ ( f^{(1)} )_\perp $
since it is orthogonal to the space of the lowest-eigenvalue functions.
It is clear that the second term 
$ \int \lambda r^2 \{ \cdots \} $
is also positive definite.
Thus we conclude that
the quadratic form (\ref{2nd order}) is positive for any $ ( f^{(1)} )_\perp $,
and deduce that
the lowest-approximated vacuum $ f^{(0)} $ is stable 
against the symmetry-breaking perturbation by $ ( f^{(1)} )_\perp $.
Hence in the vacuum $ f $ 
the orthogonal component vanishes,
i.e. $ ( f^{(1)} )_\perp = 0 $.
This implies that
the perturbed vacuum $ f = f^{(0)} + ( f^{(1)} )_\parallel $
remains 
invariant under the actions of $ H $.
The proof is over.

\section{Doublet models}
\subsection{Embedding the doublet model in the SU(2)}
Parity and charge conjugation change the sign of monopole charge
as seen above.
This fact suggests that
a model constructed from a pair of matter fields with opposite charges 
can be invariant under both $ P $ and $ C $.
Moreover, in the doublet matter model
the gauge field of the Dirac monopole 
can be embedded into an $ SU(2) $ gauge field,
which is free from the Dirac singularity.
Hence topology of the gauge field becomes trivial and 
the previous theorem that guarantees breaking of the rotational symmetry
does not apply to the doublet model.
So there is a possibility to restore the rotational symmetry 
in the doublet model.
In the following we will examine structure of the doublet model in detail.

Let us consider a doublet of scalar fields $ ( f^{q}, f^{-q} ) $.
Actually they have four components
$ ( f^{q}_+, f^{q}_-, f^{-q}_+, f^{-q}_- ) $,
where $ f_\pm $ is a smooth function 
defined in each domain $ U_\pm $.
They are related by the patching condition (\ref{patch}) as
\begin{equation}
	f^{ q}_- = e^{-2 i q \phi} f^{ q}_+, \qquad
	f^{-q}_- = e^{ 2 i q \phi} f^{-q}_+,
	\label{pairs}
\end{equation}
or equivalently, by
\begin{equation}
	\left( 
		\begin{array}{l}
		f^{q}_- \\ f^{-q}_- 
		\end{array}
	\right)
	= e^{-2 i q \sigma_3 \phi} 
	\left( 
		\begin{array}{l}
		f^{q}_+ \\ f^{-q}_+ 
		\end{array}
	\right)
	\qquad
	\mbox{in} \: U_+ \cap U_-. 
	\label{doublet}
\end{equation}
We take the same gauge field (\ref{gauge}) to define the covariant derivative
\begin{equation}
	D 
	\left( 
		\begin{array}{l} 
		f^q_\pm \\ f^{-q}_\pm 
		\end{array}
	\right)
	= d 
	\left( 
		\begin{array}{l} 
		f^q_\pm \\ f^{-q}_\pm 
		\end{array}
	\right)
	- i q A_\pm \sigma_3
	\left( 
		\begin{array}{l} 
		f^q_\pm \\ f^{-q}_\pm 
		\end{array}
	\right).
	\label{D on doublet}
\end{equation}
Let us introduce two maps $ \tau^q_\pm : U_\pm \to SU(2) $ by
\begin{equation}
	\tau^q_\pm (\theta, \phi) 
	:=
	e^{    i   \sigma_1 \pi/2}
	e^{    i q \sigma_3 \phi}
	e^{    i   \sigma_2 \theta/2}
	e^{\mp i q \sigma_3 \phi}.
	\label{transf}
\end{equation}
Then we have
\begin{equation}
	(\tau^q_-)^{-1} \cdot \tau^q_+
	=
	e^{-2 i q \sigma_3 \phi}
	\label{transf_-+}
\end{equation}
and rewrite (\ref{doublet}) as
\begin{equation}
	\left( 
		\begin{array}{l}
		f^{q}_- \\ f^{-q}_- 
		\end{array}
	\right)
	= 
	(\tau^q_-)^{-1} \cdot \tau^q_+
	\left( 
		\begin{array}{l}
		f^{q}_+ \\ f^{-q}_+ 
		\end{array}
	\right).
\end{equation}
Thus 
\begin{equation}
	\tilde{F} :=
	\tau^q_-
	\left( 
		\begin{array}{l}
		f^{q}_- \\ f^{-q}_- 
		\end{array}
	\right)
	= 
	\tau^q_+
	\left( 
		\begin{array}{l}
		f^{q}_+ \\ f^{-q}_+ 
		\end{array}
	\right)
	\label{global doublet}
\end{equation}
becomes a function that is well-defined and smooth over the whole
$ S^2 = U_+ \cup U_- $.
By substituting 
\begin{equation}
	\left( 
		\begin{array}{l}
		f^{q}_\pm \\ f^{-q}_\pm
		\end{array}
	\right)
	= 
	( \tau^q_\pm )^{-1} \tilde{F} 
	\label{global doublet inverse}
\end{equation}
into (\ref{D on doublet}) we get
\begin{equation}
	D 
	( {\tau}^q_\pm)^{-1} \tilde{F}
	=
	({\tau}^q_\pm)^{-1} 
	\Big(
		d 
		+ \tau^q_\pm d ({\tau}^q_\pm)^{-1} 
		- i q A_\pm \,
		{\tau}^q_\pm \sigma_3 ({\tau}^q_\pm)^{-1} 
	\Big) \tilde{F}
	\label{tilde D on doublet'}
\end{equation}
Then the covariant derivative of the doublet $ \tilde{F} $ is written as
\begin{equation}
	D \tilde{F} = ( d - i q \tilde{A} ) \tilde{F}
	\label{DF}
\end{equation}
with the $ SU(2) $ gauge field $ \tilde{A} $ that is defined by
\begin{eqnarray}
	-i q \tilde{A}
& := &
	 {\tau}^q_\pm d ({\tau}^q_\pm)^{-1} 
	- i q A_\pm \,
	{\tau}^q_\pm \sigma_3 ({\tau}^q_\pm)^{-1} 
	\nonumber \\
& = &
	\frac{1}{2} i \sigma_1 
	\left(
		-   \sin 2 q \phi \, d \theta 
		-2q \cos \theta  \cos 2 q \phi  \sin \theta  d \phi
	\right)
	\nonumber \\ &&
	+ \frac{1}{2} i \sigma_2 
	\left(
		    \cos 2 q \phi \, d \theta 
		-2q \cos \theta  \sin 2 q \phi  \sin \theta  d \phi
	\right)
	+ \frac{1}{2} i \sigma_3 \,
	2q \sin^2 \theta d \phi. \quad
	\label{tilde A}
\end{eqnarray}
This result is significant;
$ \tilde{A} $ is well-defined over the whole $ S^2 = U_+ \cup U_- $
and is free from the Dirac singularity.
If we introduce the orthogonal frame 
\begin{equation}
	\vect{e}^q_r :=
	\left( 
		\begin{array}{l}
		\sin \theta \, \cos 2q \phi \\
		\sin \theta \, \sin 2q \phi \\
		\cos \theta 
		\end{array}
	\right),
	\quad
	\vect{e}^q_\theta :=
	\left( 
		\begin{array}{l}
		\cos \theta \, \cos 2q \phi \\
		\cos \theta \, \sin 2q \phi \\
		-\sin \theta 
		\end{array}
	\right),
	\quad
	\vect{e}^q_\phi :=
	\left( 
		\begin{array}{r}
		- \sin 2q \phi \\
		  \cos 2q \phi \\
		0
		\end{array}
	\right),
	\label{q-frame}
\end{equation}
the resulted gauge field (\ref{tilde A}) is rewritten as
\begin{equation}
	\tilde{A}
	=
	- \frac{1}{2 q} \vect{\sigma} \cdot 
	( \vect{e}^q_\phi d \theta
	- 2q \, \vect{e}^q_\theta \sin \theta \, d \phi ).
	\label{tilde A rearranged}
\end{equation}
Thus the doublet model is reconstructed 
from the globally single-valued fields, $ \tilde{F} $ and $ \tilde{A} $.
This result manifests the mathematical fact that
the direct sum of the vector bundles of the monopole numbers $ \pm q $ 
becomes a trivial bundle.
In particular, when $ q=1/2 $, the gauge field (\ref{tilde A rearranged})
can be simplified as
\begin{equation}
	A_{\mbox{\tiny Hosotani}}
	= 
	- \frac{1}{r^2} \vect{\sigma} \cdot ( \vect{x} \times d \vect{x} ),
	\label{Hosotani field}
\end{equation}
where $ \vect{x} = (x_1,x_2,x_3) $ denotes the Cartesian coordinates.
This expression coincides with the monopole field that has been
introduced by Hosotani \cite{Hosotani S2}.
If we define a covering map of the sphere by
\begin{equation}
	\pi_{2q} : S^2 \to S^2, \qquad
	( \theta, \phi ) \mapsto ( \theta, 2q \phi ),
	\label{covering}
\end{equation}
then using the pullback by $ \pi_{2q} $ we can see that 
\begin{equation}
	2q \tilde{A} = \pi_{2q}^* A_{\mbox{\tiny Hosotani}}.
	\label{covered}
\end{equation}
The field strength accompanying with $ \tilde{A} $ is calculated to be
\begin{equation}
	\tilde{B} 
	= d \tilde{A} - i q \tilde{A} \wedge \tilde{A} 
	= - 
	\vect{\sigma} \cdot \vect{e}^q_r 
	\sin \theta d \theta \wedge d \phi.
	\label{tilde B}
\end{equation}

\subsection{Symmetries of the doublet model}
A model is defined by specifying the action or the energy functional.
We take the energy functional
\begin{equation}
        E 
	= 
        \int d \theta d \phi \, r^2 \sin \theta
        \Bigg\{
                | Df^{q} |^2 + | Df^{-q} |^2
                - \mu^2 ( |f^{q}|^2 + |f^{-q}|^2 )
                + \lambda ( |f^{q}|^2 + |f^{-q}|^2 )^2
        \Bigg\}
        \label{doublet energy}
\end{equation}
for the doublet model.
Let us examine the symmetries of this model.

{}First, 
this model is invariant under the global $ U(1) $ transformation
\begin{equation}
	\left(
		\begin{array}{l}
		f_\pm^{q} \\ f_\pm^{-q}
		\end{array}
	\right) 
	\mapsto 
	\left(
		\begin{array}{l}
		e^{-iqt} f_\pm^{q} \\ e^{iqt} f_\pm^{-q}
		\end{array}
	\right).
	\label{doublet U(1)}
\end{equation}
Second, 
it is invariant also under the $ SU(2) $ rotations of $ S^2 $,
which are given by (\ref{SU(2)f}).
Third, combined with the exchange $ f^q \leftrightarrow f^{-q} $,
parity and charge conjugation become
\begin{eqnarray}
	&& \psi_P : 
	\left(
		\begin{array}{l}
		f_\pm^{q} \\ f_\pm^{-q}
		\end{array}
	\right) 
	(\theta, \phi)
	\mapsto 
	\left(
		\begin{array}{l}
		(-1)^{(1 \mp 1) q} f^{-q}_\mp \\
		(-1)^{(1 \pm 1) q} f^q_\mp 
		\end{array}
	\right) 
	(\pi - \theta, \pi + \phi),
	\label{P doublet} 
\\
	&& \psi_C : 
	\left(
		\begin{array}{l}
		f_\pm^{q} \\ f_\pm^{-q}
		\end{array}
	\right) 
	(\theta, \phi)
	\mapsto 
	\left(
		\begin{array}{l}
		(f^{-q}_\pm)^* \\ (f^{ q}_\pm)^* 
		\end{array}
	\right) 
	(\theta, \phi).
	\label{C doublet}
\end{eqnarray}
The multiplication by phase $ (-1)^q $ is included in $ \psi_P $ to ensure that
$ (\psi_P)^2 = $ identity.
Then the energy functional (\ref{doublet energy}) is invariant under both
$ P $ and $ C $.
For later reference, we write down the $ CP $ transformation;
\begin{equation}
	\psi_{CP} = \psi_C \circ  \psi_P : 
	\left(
		\begin{array}{l}
		f_\pm^{q} \\ f_\pm^{-q}
		\end{array}
	\right) 
	(\theta, \phi)
	\mapsto 
	\left(
		\begin{array}{l}
		(-1)^{(1 \pm 1) q} (f^{ q}_\mp)^* \\
		(-1)^{(1 \mp 1) q} (f^{-q}_\mp)^* 
		\end{array}
	\right) 
	(\pi - \theta, \pi + \phi).
	\label{CP doublet} 
\end{equation}
Fourth, our model has another $ SU(2) $ symmetry
accompanying with the doublet structure.
If we regard the doublet
\begin{equation}
	\left( 
	\begin{array}{l} f^q \\ (f^{-q})^* \end{array}
	\right)
	\label{qq doublet}
\end{equation}
as independent variables, the both fields have the same charge $ q $,
and therefore the doublet admits the $ SU(2) $ transformations,
under which the energy functional (\ref{doublet energy}) is left invariant.
This symmetry is denoted by $ SU(2)_f $ to remind us of the flavor symmetry.

It should be noted that the $ \psi_C $ transformation 
is not an independent symmetry
as seen in the following:
when it acts on the doublet (\ref{qq doublet}),
(\ref{C doublet}) is equivalently written as
\begin{equation}
	\psi_C : 
	\left(
		\begin{array}{l}
		f_\pm^{q} \\ (f_\pm^{-q})^*
		\end{array}
	\right) 
	(\theta, \phi)
	\mapsto 
	\left(
		\begin{array}{l}
		(f^{-q}_\pm)^* \\ f^{q}_\pm 
		\end{array}
	\right) 
	(\theta, \phi)
	\label{C' doublet}
\end{equation}
But this is nothing but the transformation
\begin{equation}
	\psi_C = \sigma_1 
	= i \, e^{-i \sigma_1 \pi/2 }
	= e^{i \pi/2} \, e^{-i \sigma_1 \pi/2 },
	\label{C is in U(1)xSU(2)}
\end{equation}
which is an element of $ U(1) \times SU(2)_f $.
Thus we conclude that 
the symmetries of the doublet model (\ref{doublet energy}) are
\begin{equation}
	G_{\mbox{\tiny doublet}} = U(1) \times SU(2) \times SU(2)_f \times P.
	\label{doublet symmetry}
\end{equation}

\subsection{q=1/2 doublet}
{}For the doublet model of $ q=1/2 $
we can find an exact vacuum configuration, which realizes the lowest energy of
the functional (\ref{doublet energy}).
The exact vacuum is given by
\begin{equation}
	\left( 
		\begin{array}{l} 
		f^{1/2}_\pm \\ (f^{-1/2}_\pm)^* 
		\end{array}
	\right)
	(\theta,\phi) 
	=
	v
	\left( 
		\begin{array}{l}
		D^{1/2}_{1/2,1/2}  \\
		D^{1/2}_{-1/2,1/2} 
		\end{array}
	\right)
	(\theta,\phi,\mp \phi) 
	=
	v e^{\pm i \phi/2}
	\left( 
		\begin{array}{l}
		e^{- i \phi/2} \cos (\theta/2) \\
		e^{  i \phi/2} \sin (\theta/2) 
		\end{array}
	\right).
	\label{1/2 doublet}
\end{equation}
Since these are the eigenfunctions for the lowest eigenvalue 
of the monopole Laplacian,
the configuration (\ref{1/2 doublet}) realizes 
the lowest value of the kinetic energy term 
in (\ref{doublet energy}).
Moreover, since $ |f^{1/2}|^2 + |f^{-1/2}|^2 = v^2 $ is constant,
(\ref{1/2 doublet}) can realize 
the lowest value of the potential energy term in (\ref{doublet energy}).
Thus (\ref{1/2 doublet}) is the exactly lowest energy state.
Then the total energy (\ref{doublet energy}) is evaluated as
\begin{equation}
        E 
	= 
        4 \pi
        \Bigg\{
                \bigg( \frac{1}{2} - \mu^2 r^2 \bigg) v^2
                + \lambda r^2 v^4 
        \Bigg\},
        \label{doublet energy for 1/2}
\end{equation}
which is minimized by
\begin{equation}
	v^2 = 
	\left\{
		\begin{array}{ll}
		    0 
		    & \quad \mbox{for} \quad r \leq \sqrt{1/2} \, \mu^{-1}, 
		    \\
		    \Big( 1 - \dfrac{1}{2 \mu^2 r^2 } \Big) 
		    \dfrac{\mu^2}{2 \lambda} 
		    & \quad \mbox{for} \quad r > \sqrt{1/2} \, \mu^{-1}.
		\end{array} 
	\right.
        \label{v^2 for 1/2}
\end{equation}

By applying the embedding map (\ref{global doublet}) 
we can rewrite the doublet solution (\ref{1/2 doublet}) in a global form as
\begin{eqnarray}
	\tilde{F} 
& = &
	\tau^{1/2}_\pm 
	\left( 
		\begin{array}{l}
		f^{1/2}_\pm \\ f^{-1/2}_\pm 
		\end{array}
	\right)
	\nonumber \\
& = &
	v \,
	e^{    i \sigma_1 \pi/2}
	e^{    i \sigma_3 \phi/2}
	e^{    i \sigma_2 \theta/2}
	e^{\mp i \sigma_3 \phi/2}
	\left( 
		\begin{array}{l}
		e^{\pm i \phi/2}
		e^{- i \phi/2} \cos (\theta/2) \\
		e^{\mp i \phi/2}
		e^{- i \phi/2} \sin (\theta/2) 
		\end{array}
	\right)
	\nonumber \\
& = &
	v 
	\left( 
		\begin{array}{l}
		0 \\ i
		\end{array}
	\right).
\end{eqnarray}
Thus the field $ \tilde{F} $ is reduced to a constant function
and has no vortices in $ S^2 $.
Hence it is naturally expected that the vacuum is invariant 
under the rotations.

Let us examine the preserved symmetry of the vacuum (\ref{1/2 doublet}).
{}For $ q=1/2 $, parity (\ref{P doublet}) becomes
\begin{equation}
	\psi_P : 
	\left( 
		\begin{array}{l}
		f^{1/2}_\pm \\ f^{-1/2}_\pm 
		\end{array}
	\right)
	\mapsto 
	\left( 
		\begin{array}{l}
		\pm f^{-1/2}_\mp \\
		\mp f^{1/2}_\mp 
		\end{array}
	\right)
	(\pi - \theta, \pi + \phi),
	\label{P doublet 1/2} 
\end{equation}
and it can be easily seen that 
the vacuum (\ref{1/2 doublet}) is invariant under $ \psi_P $.
Apparently, the vacuum is not invariant
under either the $ SU(2) $ rotations or the $ SU(2)_f $ flavor transformations.
However, it can be shown that the vacuum is actually invariant under 
actions of the diagonal elements of $ SU(2) \times SU(2)_f $,
which are expressed as
\begin{equation}
	\psi_g :
	\left( 
		\begin{array}{l} 
		f^{1/2}_\alpha \\ (f^{-1/2}_\alpha)^* 
		\end{array}
	\right) (p)
	\mapsto
	e^{i \omega_{\alpha \beta} (g; p)/2} \, g
	\left( 
		\begin{array}{l} 
		f^{1/2}_\beta \\ (f^{-1/2}_\beta)^* 
		\end{array}
	\right) (g^{-1}p),
	\qquad
	g \in SU(2).
	\label{diagonal SU(2)xSU(2)}
\end{equation}
We call this symmetry the flavored rotational symmetry 
because 
it transforms the flavor doublet 
and simultaneously rotates the point $ p $ in the sphere.
We will give the proof of the invariance later in a more general context.
These results are summarized as follows:
the vacuum of the $ q=1/2 $ doublet model has the symmetry
\begin{equation}
	H_{\mbox{\tiny 1/2 doublet}} = SU(2)' \times P.
	\label{broken doublet symmetry}
\end{equation}
It is to be noted that the $ C $ symmetry (\ref{C doublet}) 
is broken spontaneously in this model.

\section{Multiplet models}
In this section
we discuss general aspects of models that have several matter fields
in the monopole background field.

\subsection{Disentanglement of the Dirac singularity}
Remember that
the domains $ U_\pm $ of the sphere $ S^2 $ are defined as
$ U_\pm = \{ (\theta, \phi) \in S^2 | \cos \theta \ne \mp 1 \} $.
Generally we can construct multiplets of scalar fields
\begin{equation}
	F_\pm =
	\left( 
		\begin{array}{c} 
		f^{q_1}_\pm \\ 
		f^{q_2}_\pm \\ 
		\vdots \\
		f^{q_n}_\pm 
		\end{array}
	\right),
	\label{multiplet}
\end{equation}
where $ \{ 2 q_i \}_{i=1, \cdots, n} $ are integers.
Each multiplet $ F_\pm $ is a single-valued continuous function 
in each domain $ U_\pm $
and they are patched by the gauge transformation
\begin{equation}
	F_- =
	T(\phi) F_+ = 
	\left( 
		\begin{array}{llll}
		e^{-2 i q_1 \phi} & 0 & \cdots & 0 \\
		0 & e^{-2 i q_2 \phi} & & 0 \\ 
		\vdots & & \ddots & \vdots \\
		0 & 0 & \cdots & e^{-2 i q_n \phi} \\ 
		\end{array}
	\right)
	F_+
	\label{patched multiplet}
\end{equation}
in $ U_+ \cap U_- $.
If we put the diagonal matrix
$ Q = \mbox{diag} ( q_1, q_2, \cdots, q_n ) $,
we may write $ T(\phi) = e^{-2 i Q \phi} $.
Then the covariant derivative of $ F_\pm $ is defined by
\begin{equation}
	D F_\pm
	=
	( d - i Q A_\pm ) F_\pm
	\label{D on F}
\end{equation}
with the monopole gauge field 
$ A_\pm = ( \pm 1 - \cos \theta ) d \phi $.
We may replace $ f^{q_i} $ by its conjugate field $ (f^{q_i})^* $,
which has a monopole charge $ - q_i $.
Note that the fields
$ (F_+, A_+) $ are ill-defined at the south pole, $ \theta = \pi $, 
of the sphere,
whereas
$ (F_-, A_-) $ are ill-defined at the north pole, $ \theta = 0 $.
In the previous section,
we studied the doublet model with $ (q_1, q_2) = (q,-q) $ 
and rewrote the model in terms of globally well-defined fields,
which are free from singularities,
by embedding the gauge field in the $ SU(2) $ group.
This argument is applicable to general models with larger multiplets.
We can prove that if 
\begin{equation}
	\sum_{i=1}^n q_i = 0,
	\label{disentanglement condition}
\end{equation}
then the monopole gauge field is embedded in the $ SU(n) $ group,
and that the fields are transformed into globally singled-valued ones.

Let us prove the above statement.
The matrix-valued function $ T(\phi) $ in (\ref{patched multiplet}) is a map
$ T : S^1 \to U(n) $.
If the condition (\ref{disentanglement condition}) is satisfied,
$ T(\phi) $ is a continuous map $ T : S^1 \to SU(n) $.
Because the first homotopy group of $ SU(n) $ is trivial, 
namely $ \pi_1 (SU(n)) = 0 $,
there exists a continuous map
\begin{equation}
	T_+ : [0, \pi] \times S^1 \to SU(n);
	\qquad
	(\theta, \phi) \mapsto T_+ (\theta, \phi) 
	\label{T_+}
\end{equation}
such that 
$ T_+ (0, \phi) = 1 $ and $ T_+ (\pi, \phi) = T(\phi) $.
Now we let $ T_+ $ operate on $ F_+ $ and define
\begin{equation}
	\tilde{F} = T_+ F_+ : [0, \pi] \times S^1 \to \C^n.
	\label{global Phi}
\end{equation}
Then we have
$ \tilde{F} (0,   \phi) = F_+ (0,   \phi) $ and
$ \tilde{F} (\pi, \phi) = T(\phi) F_+(\pi, \phi) = F_- (\pi, \phi) $,
and hence 
$ \tilde{F} $ becomes a single-valued continuous function
$ \tilde{F} : S^2 \to \C^n $.
The covariant derivative of $ F_+ $ (\ref{D on F}) implies
\begin{equation}
	D F_+
	=
	( d - i Q A_+ ) F_+
	=
	( d - i Q A_+ ) T_+^{-1} \tilde{F}
	=
	T_+^{-1} 
	\Big(
		d 
		+ T_+ d T_+^{-1}
		- i T_+ Q T_+^{-1} A_+
	\Big) \tilde{F}
	\label{D on Phi}
\end{equation}
Hence the covariant derivative of the multiplet $ \tilde{F} $ is written as
\begin{equation}
	D \tilde{F} = ( d - i \tilde{A} ) \tilde{F}
	\label{D Phi}
\end{equation}
with the $ SU(n) $ gauge field $ \tilde{A} $ that is defined by
\begin{equation}
	\tilde{A}
	=
	i T_+ d T_+^{-1} + 
	T_+ Q T_+^{-1} A_+.
	\label{global SU(n)}
\end{equation}
Of course, $ \tilde{A} $ is well-defined over $ U_+ $.
In the limit $ \theta \to \pi $, 
we have $ T_+ \to T = e^{-2 i Q \phi} $ and 
\begin{equation}
	\tilde{A}
	\to
	T Q T^{-1} A_+ + i T d T^{-1}
	= Q ( A_+ - 2 d \phi) 
	= Q A_-,
	\label{continuity}
\end{equation}
therefore, $ \tilde{A} $ is well-defined over the whole $ S^2 $.
Thus we arrive at the consequence that
the fields 
$ ( F_\pm, A_\pm ) $
are transformed to the globally defined fields
$ ( \tilde{F}, \tilde{A} ) $.

To illustrate the above theorem we may consider a multiplet of charges
$ ( q_1, q_2, \cdots , $ $ q_{2j}, q_{2j+1} )
= (j, j-1, \cdots, -j+1, -j ) $
for an integer or half-integer $ j $.
Then the matrix $ Q $ is identified with the generator $ J_3 $ of $ SU(2) $
in the spin $ j $ representation.
If we take
\begin{equation}
	T_+ (\theta, \phi) = 
	e^{i J_2 \theta} e^{i J_3 \phi} e^{-i J_2 \theta} e^{-i J_3 \phi},
	\label{2j+1 transformation}
\end{equation}
it has the desired properties;
$ T_+ (0, \phi) = 1 $ and $ T_+ (\pi, \phi) = e^{-2 i Q \phi} $.
In this case,
the gauge field $ \tilde{A} $ is embedded in the subgroup
$ SU(2) \subset SU(n) $.

\subsection{Exact solutions}
In the subsection 3.3 
we obtained the exact vacuum solution of the $ q = 1/2 $ doublet model.
A similar method can be used 
to obtain a series of exact solutions of larger multiplet models
as shown below.
Let us consider a multiplet $ F $ of fields with degenerated charges
$ (q_1, q_2, \cdots , q_{2j}, q_{2j+1} )
= (j,j, \cdots, j,j ) $ 
for a fixed $ j > 0 $
and define the model by the energy functional
\begin{equation}
        E = 
        \int d \theta d \phi \, r^2 \sin \theta
        \Big\{
                D F^\dagger D F 
                - \mu^2     F^\dagger F 
                + \lambda ( F^\dagger F )^2
        \Big\}.
        \label{multiplet energy}
\end{equation}
Apparently, this energy is invariant under the flavor transformations,
\begin{equation}
	F_\pm \mapsto F'_\pm = T F_\pm, \qquad T \in SU(2j+1).
	\label{flavor}
\end{equation}
This symmetry is denoted by $ SU(2j+1)_f $.
The energy functional is invariant also under the $ CP $ transformation
\begin{equation}
	\varphi_{CP} :
	F_\pm (\theta, \phi) \mapsto
	e^{\mp i j \pi} 
	F_\mp^* (\pi-\theta, \pi+\phi),
	\label{CP multiplet}
\end{equation}
whose definition is brought from (\ref{CP f}).
Thus the model has the symmetry
\begin{equation}
	G_{\mbox{\tiny multiplet}} = 
	U(1) \times SU(2) \times SU(2j+1)_f \times CP.
	\label{multiplet symmetry}
\end{equation}

We can immediately write down the exact lowest energy state as
\begin{equation}
	F_\pm (\theta,\phi)
	=
	\left( 
		\begin{array}{l} 
		f_{j}    \\
		f_{j-1}  \\
		\vdots       \\
		f_{-j+1} \\
		f_{-j} 
		\end{array}
	\right)_\pm \!\!\! (\theta,\phi)
	= v
	\left( 
		\begin{array}{l} 
		D^{j}_{j,j}     \\
		D^{j}_{j-1,j}  \\
		\vdots         \\
		D^{j}_{-j+1,j} \\
		D^{j}_{-j,j}
		\end{array}
	\right) (\theta,\phi,\mp \phi),
	\label{multiplet solution}
\end{equation}
which is a natural generalization of the doublet solution (\ref{1/2 doublet})
to the larger multiplet.
To see that it minimizes the energy,
first, remember that the matrix elements $ \{ D^j_{m,j} \} $ are 
eigenfunctions belonging to the lowest eigenvalue of the monopole Laplacian.
Hence they minimize the kinetic term in (\ref{multiplet energy}).
Second, note that
the squared norm $ F^\dagger F = v^2 $ is constant
for (\ref{multiplet solution}).
Hence it minimizes the potential term in (\ref{multiplet energy})
by a suitable choice of $ v $.
By substituting (\ref{multiplet solution}) into (\ref{multiplet energy})
we evaluate the energy as
\begin{equation}
	E = 
	4 \pi
        \Bigg\{
                \bigg( j - \mu^2 r^2 \bigg) v^2
                + \lambda r^2 v^4 
        \Bigg\},
        \label{multiplet energy estimated}
\end{equation}
which is minimized by
\begin{equation}
	v^2 = 
	\left\{
		\begin{array}{ll}
		    0 
		    & \quad \mbox{for} \quad r \leq \sqrt{j} \, \mu^{-1}, 
		    \\
		    \Big( 1 - \dfrac{j}{\mu^2 r^2 } \Big) 
		    \dfrac{\mu^2}{2 \lambda} 
		    & \quad \mbox{for} \quad r > \sqrt{j} \, \mu^{-1}.
		\end{array} 
	\right.
        \label{v^2 for multiplet}
\end{equation}

A remaining question is to ask the symmetry of the vacuum.
The answer is that the vacuum (\ref{multiplet solution})
has the symmetry
\begin{equation}
	H_{\mbox{\tiny multiplet}} = SU(2)' \times CP'.
	\label{broken multiplet symmetry}
\end{equation}
In the following we identify the symmetry $ H_{\mbox{\tiny multiplet}} $.
We write the representation matrix element of $ SU(2) $ as 
$ D^j_{mm'} (g) = \bra j,m | g | j,m' \ket $.
Then the component field of (\ref{multiplet solution}) is written as
\begin{equation}
	(f_m)_\pm (p) = v D^j_{mj} (s_\pm(p))
	\label{f pm}
\end{equation}
with use of the maps $ s_\pm : U_\pm \to SU(2) $ defined in (\ref{sections}).
Under the action of $ g \in SU(2) $, it is transformed to
\begin{eqnarray}
	(f_m)_\beta (g^{-1} p) 
& = &
	v D^j_{mj} (s_\beta(g^{-1} p))
	\nonumber \\
& = &
	v D^j_{mj} ( g^{-1} \cdot s_\alpha(p) \cdot 
	s_\alpha(p)^{-1} g  s_\beta(g^{-1} p))
	\nonumber \\
& = &
	v \sum_{m',q'}
	D^j_{mm'} ( g^{-1} ) \,
	D^j_{m'q'} ( s_\alpha(p) ) \,
	D^j_{q'j} ( s_\alpha(p)^{-1} g  s_\beta(g^{-1} p) ),
	\label{f pm g}
\end{eqnarray}
where $ \alpha $ and $ \beta $ denote $ + $ or $ - $.
By the definition of the Wigner rotation,
namely by (\ref{Wigner}) and (\ref{Wigner in U(1)}),
we get
\begin{equation}
	D^j_{q'j} ( s_\alpha(p)^{-1} g  s_\beta(g^{-1} p) )
	=
	\bra j,q' | W_{\alpha \beta} (g;p) | j,j \ket
	=
	\bra j,q' | e^{-i J_3 \omega_{\alpha \beta}} | j,j \ket
	=
	\delta_{q'j} \, e^{-i j \omega_{\alpha \beta}}.
	\label{Wigner again}
\end{equation}
{}From (\ref{f pm}), (\ref{f pm g}) and (\ref{Wigner again})
we deduce 
\begin{equation}
	(f_{m})_\beta (g^{-1} p) 
	=
	\sum_{m'}
	D^j_{mm'} ( g^{-1} ) \,
	(f_{m'})_\alpha (p)   \,
	e^{-i j \omega_{\alpha \beta}},
	\label{f pm SU(2) action}
\end{equation}
or
\begin{equation}
	e^{i j \omega_{\alpha \beta}}
	\sum_{m'}
	D^j_{mm'} ( g )
	(f_{m'})_\beta (g^{-1} p) 
	=
	(f_{m})_\alpha (p),
	\label{f pm invariant}
\end{equation}
Hence the configuration (\ref{multiplet solution}) is invariant
under the combined transformation
by the rotation $ g \in SU(2) $ 
with the flavor transformation $ D^j_{mm'} (g) \in SU(2j+1)_f $. 
The elements $ \{ D^j_{mm'} (g) \} $
forms a subgroup $ SU(2)_f \subset SU(2j+1)_f $.
Namely, the diagonal part
$ SU(2)' = $ diag $ ( SU(2) \times SU(2)_f ) $
is identified as the symmetry group of the vacuum.
In particular, 
when $ j =1/2 $,
the diagonal transformation (\ref{f pm invariant}) realizes
the transformation (\ref{diagonal SU(2)xSU(2)}) 
of the doublet as mentioned before.
We call $ SU(2)' $ the flavored rotational symmetry again.

{}From the definition (\ref{sections}) we deduce that
\begin{eqnarray}
        s_\pm (\theta, \phi) e^{i \sigma_1 \pi/2}
& = &
	e^{-i \sigma_3 \phi  /2}
	e^{-i \sigma_2 \theta/2}
	e^{\pm i \sigma_3 \phi  /2} 
	\cdot i \sigma_1
	\nonumber \\
& = &
	e^{-i \sigma_3 \phi  /2}
	\cdot i \sigma_1 \cdot
	e^{ i \sigma_2 \theta/2}
	e^{\mp i \sigma_3 \phi  /2} 
	\nonumber \\
& = &
	e^{-i \sigma_3 \phi  /2}
	( -i \sigma_3 )
	( -i \sigma_2 )
	e^{ i \sigma_2 \theta/2}
	e^{\mp i \sigma_3 \phi  /2} 
	\nonumber \\
& = &
	e^{-i \sigma_3 \phi  /2}
	e^{-i \sigma_3 \pi/2}
	e^{-i \sigma_2 \pi/2}
	e^{ i \sigma_2 \theta/2}
	e^{\mp i \sigma_3 \phi  /2} 
	\nonumber \\
& = &
	s_\mp (\pi - \theta, \pi + \phi) 
	e^{\pm i \sigma_3 \pi /2},
        \label{P on sections}
\end{eqnarray}
namely,
\begin{eqnarray}
        s_\pm (\theta, \phi) 
& = &
	s_\mp (\pi - \theta, \pi + \phi) 
	e^{\pm i \sigma_3 \pi /2}
	e^{-i \sigma_1 \pi/2}.
        \label{cal}
\end{eqnarray}
Hence, we have
\begin{equation}
        D^j_{mj}
        ( s_\pm (\theta, \phi) )
	=
	\sum_{m', \, q'}
	D^j_{m m'} ( s_\mp (\pi - \theta, \pi + \phi) )
	D^j_{m'q'} ( e^{\pm i \sigma_3 \pi /2} )
	D^j_{q'j } ( e^{-i \sigma_1 \pi/2} ).
	\label{Parity of section}
\end{equation}
It is an easy exercise to verify that 
the representation matrix elements of $ SU(2) $ satisfy
\begin{eqnarray}
&&	D^j_{m m'} ( s_\mp (\pi - \theta, \pi + \phi) )
	=
	e^{i \pi (m - m')}
	(D^j_{-m, -m'})^* ( s_\mp (\pi - \theta, \pi + \phi) ),
	\label{D*} \\
&&	D^j_{m'q'} ( e^{\pm i \sigma_3 \pi /2} )
	=
	\delta_{m'q'} \, e^{\pm i q' \pi },
        \label{i sigma_3} \\
&&	D^j_{q'j} ( e^{-i \sigma_1 \pi/2} )
	=
	\delta_{q',-j} \, e^{-i j \pi}.
	\label{i sigma_1} 
\end{eqnarray}
By substituting them into (\ref{Parity of section}), we get
\begin{equation}
        D^j_{mj} ( s_\pm (\theta, \phi) )
        =
	e^{i m \pi}
	e^{\mp i j \pi }
	(D^j_{-m, j})^* ( s_\mp (\pi - \theta, \pi + \phi) ).
	\label{CP property of matrix}
\end{equation}
By the definition of $ CP $ in (\ref{CP multiplet}),
the above equation can be written as
\begin{equation}
        ( D^j_{mj} \circ s_\pm ) (\theta, \phi) 
        =
	e^{i m \pi}
	\varphi_{CP} ( D^j_{-m, j} \circ s_\pm ) (\theta, \phi ).
	\label{CP of multiplet}
\end{equation}
Therefore, if we define the flavor conjugate transformation by
\begin{equation}
	\varphi_F : 
	F_\pm = 
	\left( 
		\begin{array}{l} 
		f_{j}    \\
		f_{j-1}  \\
		\vdots       \\
		f_{-j+1} \\
		f_{-j} 
		\end{array}
	\right)_\pm 
	\mapsto \;
	F'_\pm = 
	\left( 
		\begin{array}{l} 
		e^{i j \pi} f_{-j}    \\
		e^{i (j-1) \pi} f_{-j+1}  \\
		\vdots       \\
		e^{-i (j-1) \pi} f_{j-1} \\
		e^{-i j \pi} f_{j} 
		\end{array}
	\right)_\pm ,
	\label{flavor parity}
\end{equation}
which is an element of the group $ U(1) \times SU(2j+1)_f $, then we have
\begin{equation}
        ( D^j_{mj} \circ s_\pm ) (\theta, \phi) 
        =
	( \varphi_F \circ \varphi_{CP} )
	( D^j_{mj} \circ s_\pm ) (\theta, \phi ).
	\label{F CP}
\end{equation}
Thus we see that
the vacuum configuration (\ref{f pm}) is invariant under
the transformation $ \varphi_F \circ \varphi_{CP} $. 
This symmetry is denoted by $ CP' $.
Thus we accomplish identification of the symmetry of the vacuum
as $ H_{\mbox{\tiny multiplet}} = SU(2)' \times CP' $.

This result is remarkable
in comparison with the result of an ordinary model in the flat space-time.
Let us consider
a model that consists of a multiplet of scalar fields
$ F(x) = ( f_j(x), \cdots, f_{-j}(x) ) $ in the flat space-time.
Suppose that 
the model has the global $ SU(2j+1) $ symmetry 
under which $ F(x) $ obeys the fundamental representation of $ SU(2j+1) $,
and that the fields develop nonzero vacuum expectation values.
If the translational symmetry is preserved,
the expectation value $ \bra F(x) \ket $ is 
independent of the space-time point $ x $,
and hence the symmetry is broken from $ SU(2j+1) $ to $ SU(2j) $.
This is an ordinary pattern of symmetry breaking 
caused by the fundamental Higgs field.
But, in our model in $ S^2 $, 
the one fundamental scalar multiplet breaks 
the rotation-flavor symmetry from $ SU(2) \times SU(2j+1) $ to $ SU(2)' $.
It may be worthwhile to remember that
this peculiar patter of symmetry breaking is caused by the monopole gauge field.

\subsection{Phase structures of generic models}
In regard to a general model that consists of the multiplet
(\ref{multiplet}) with an arbitrary charge pattern $ (q_1, q_2, \cdots, q_n ) $,
we do not yet have a complete consequence on its phase structure.
However, we can deduce the following result confidently.
In such a general model,
each field $ f_i $ has different mass parameter $ \mu_i $
and different critical radius $ r_i = \sqrt{q_i}/\mu_i $.
Let us set $ r_s = $ min $ \{ r_i \}_{i=1, \cdots, n} $.
We can say that
when the radius $ r $ of $ S^2 $ is smaller than $ r_s $, namely $ r < r_s $, 
all the fields have vanishing vacuum expectation values
and hence the symmetry is not broken.
When $ r > r_s $, the field $ f_s $ begins to develop 
a nonzero vacuum expectation value,
and the symmetry is broken.

If the model has an increasing sequence of critical radii
$ r_1 < r_2 < \cdots $,
it may exhibit a pattern of symmetry breakings 
that develops step by step when the radius of the sphere increases.
But the detail of the symmetry pattern depends on the other parameters 
of the model.
At the present stage we do not have a consequence applicable to such
a general class of models.

\section{Conclusion}
We have studied
the scalar field in the monopole background in $ \R^n \times S^2 $
with the charge $ q =1/2, 1, 3/2, \cdots $.
We showed that
the field develops $ 2|q| $ vortices
when the radius of $ S^2 $ is larger 
than the critical radius $ r_q = \sqrt{|q|}/\mu $.
Then the rotational symmetry of $ S^2 $ is broken.
The vortices repel each other 
and settle down at the furthest separated points on $ S^2 $.
We found that in the doublet model with the charge $ (q, -q) $
the vortices do not appear and the modified rotational symmetry is left unbroken.
We also 
obtained the exact vacuum configuration of the model with the multiplet
$ (q_1, q_2, \cdots , q_{2j}, q_{2j+1} ) = (j,j, \cdots, j,j ) $
and found that 
the rotation-flavor symmetry $ SU(2) \times SU(2j+1) $ 
is spontaneously broken 
to the flavored rotational symmetry $ SU(2)' $.
We classified the patters of symmetry breaking of these models.

We would like to put remarks on further developments of this work.
The two-sphere, which is the extra dimensions of the present model, 
is a kind of homogeneous space $ S^2 = SU(2)/U(1) $. 
We have seen that
the model defined in $ S^2 $ has 
the $ SU(2) $ symmetry and the $ U(1) $ gauge field.
As one of the directions for developments,
we can construct models 
in higher dimensional manifolds than $ S^2 $,
which have larger global symmetries than $ SU(2) $
and larger gauge symmetries than $ U(1) $.
For example, the $ n $-dimensional sphere $ S^n $ is also a
homogeneous space, which is written as $ S^n = SO(n+1)/SO(n) $.
More generally, a homogeneous space $ M = G/H $ induces 
a gauge field associated with the nonabelian gauge group $ H $
as has been shown in \cite{Tanimura}.
The model defined in $ G/H $ can 
have various topological defects other than vortices
and
exhibit more complicated patterns of symmetry breaking.

It is strongly desirable to include fermions in our formulation
for application to realistic models.
Fermions open ways for further development:
chiral fermions, the generation structure,
supersymmetry, its breaking, flavor mixing, and so on.
When there is a topological defect in extra dimensions,
zero modes of the Dirac operator are trapped in the defect
and become chiral fermions in the four dimensions.
Then the number of fermion generations
coincides with the topological number of the defect.
%
Our analysis indicates that a degenerated vortex 
with a large topological number
is unstable and decays into separated vortices.
Thus 
we need 
more careful analyses on dynamics of topological defects 
to build more realistic models.
In particular, the back reaction of the gauge field,
which is known as the Meissner effect, should be considered.

In this paper 
we assume the existence of the monopole background in the extra dimensions.
We also assume the negative square mass $ - \mu^2 $ of the scalar field.
Of course, their origins are to be pursued further.
We would like to mention studies by other people, 
which are related to our problem.
Hosotani \cite{Hosotani S2} showed that
when fermions live in the space-time $ \R^n \times S^2 $,
the vacuum energy is lowered by the monopole background.
Accordingly, the monopole can be generated dynamically.
Moreover, Randjbar-Daemi, Salam and Strathdee \cite{Salam:Maxwell}
proved that
the monopole solutions of the classical Einstein-Maxwell theory
are stable.
On the other hand, 
the same authors \cite{Salam:YM}
showed that
a solution of the classical Einstein-Yang-Mills theory in $ \R^4 \times S^2 $
is unstable
if the solution is invariant under the rotations of $ S^2 $.
Although these models do not directly solve our problem about 
the origins of the monopole background and of the tachyonic mode, 
they give insights about dynamics in higher dimensions.

Moreover,
finding a new mechanism of supersymmetry breaking \cite{Takenaga}
is the original motivation of this work.
To realize it we need to equip fermions in the extra dimensions 
in a supersymmetric manner.
Furthermore, the twisted boundary condition could provide
a new mechanism of the flavor mixing;
if the eigenstates of the boundary condition differ from 
the eigenstates of the interaction,
then the mixing occurs.
We call this phenomenon the topological mixing mechanism
and leave it for future study.
Finally, it is an interesting question
to ask cosmological implications of the critical radius.


\section*{Acknowledgments}
The authors wish to thank 
N. Haba,
H. Hatanaka, 
M. Hayakawa, 
Y. Hosotani, 
H. Ikemori,
S. Iso, 
T. Kugo, 
C.S. Lim, 
S. Matsumoto, 
Y. Nagatani, 
M. Nakahara, 
H. Nakano, 
K. Ohnishi, 
H. Otsu, 
N. Sakai, 
H. So,
M. Tachibana, 
K. Takenaga,
I. Tsutsui
and
K. Yamawaki
for valuable comments and criticisms.
This work was supported by Grant-in-Aids for Scientific Research 
({\#}12640275 for M.S. and {\#}12047216 for S.T.)
from
Ministry of Education, Culture, Sports, Science and Technology of Japan.


\renewcommand{\thesection}{Appendix A.}
\renewcommand{\theequation}{A.\arabic{equation}}


\baselineskip 5mm 

\end{document}